\documentclass[a4paper,11pt]{article}
\pdfoutput=1 % if your are submitting a pdflatex (i.e. if you have
\usepackage{jheppub} % for details on the use of the package, please
\usepackage[T1]{fontenc} % if needed
\usepackage{subcaption}
\preprint{PUPT-2604}

\title{Simplicity of AdS Supergravity at One Loop}

\author[]{Luis F. Alday$^{a}$,}
\affiliation[]{$^{a}$Mathematical Institute, University of Oxford,
Andrew Wiles Building, Radcliffe Observatory Quarter, Woodstock Road, Oxford, OX2 6GG, U.K.}
\author[]{Xinan Zhou$^{b}$}
\affiliation[]{$^{b}$Princeton Center for Theoretical Science, Princeton University, Princeton, NJ 08544, USA }

\emailAdd{alday@maths.ox.ac.uk}
\emailAdd{xinanz@princeton.edu}

\abstract{We demonstrate the simplicity of $AdS_5\times S^5$ IIB supergravity at one loop level, by studying non-planar holographic four-point correlators in Mellin space. We develop a systematic algorithm for constructing one-loop Mellin amplitudes from the tree-level data, and obtain a simple closed form answer for the $\langle \mathcal{O}_2^{SG}\mathcal{O}_2^{SG}\mathcal{O}_p^{SG}\mathcal{O}_p^{SG} \rangle$ correlators. The structure of this expression is remarkably simple, containing only simultaneous poles in the Mellin variables. We also study the flat space limit of the Mellin amplitudes, which reproduces precisely  the IIB supergravity one-loop amplitude in ten dimensions. Our results  provide nontrivial evidence for the persistence of the hidden conformal symmetry at one loop.}

\begin{document}
\maketitle
\flushbottom
\section{Introduction}
Studying scattering amplitudes of weakly coupled theories in flat space has revealed many enticing mathematical structures (see, {\it e.g.}, \cite{Elvang:2015rqa,nima} for recent textbook presentations). Many of these structures are ``hidden'' as they are invisible from the Lagrangian formulation of the theory. For this reason, the unexpected structures in the amplitudes provide rare opportunities to understand the theory from a different perspective. One may imagine to extend the scattering amplitude program in flat space to include amplitudes on a curved background. The prime choice of the background would be the Anti de Sitter space because of the  AdS/CFT duality. Via the correspondence, the scattering amplitudes in the bulk are mapped to the conformal correlation functions on the boundary. The additional conformal symmetry offers us extra leverage, and we can exploit the modern methods of the conformal bootstrap.

While the program of AdS scattering amplitudes (or holographic correlators) was initiated a long time ago, only recently have truly efficient computational methods been developed. In the paradigmatic example of IIB supergravity on $AdS_5\times S^5$, which is dual to the infinite 't Hooft coupling limit of the 4d $\mathcal{N}=4$ SYM theory, only a handful of explicit results of four-point functions had been computed in the past \cite{DHoker:1999kzh,Arutyunov:2000py,Arutyunov:2003ae,Arutyunov:2002fh,Dolan:2006ec,Berdichevsky:2007xd,Uruchurtu:2008kp,Uruchurtu:2011wh}. It turns out that one should instead consider the holographic correlator as a whole, which is much simpler and more rigid than individual diagrams. In \cite{Rastelli:2016nze,Rastelli:2017udc}, methods inspired by the conformal bootstrap were introduced, which translates the task of computing correlation functions into solving an algebraic bootstrap problem in Mellin space. The new approaches have led to a simple compelling conjecture for {\it all} one-half BPS four-point functions in the tree-level supergravity limit \cite{Rastelli:2016nze,Rastelli:2017udc}, which was later verified by explicit calculations in a large number of examples \cite{Arutyunov:2017dti,Arutyunov:2018neq,Arutyunov:2018tvn}. These techniques have also been generalized and combined with other methods to produce new results in other backgrounds \cite{Zhou:2017zaw,Rastelli:2017ymc,Zhou:2018ofp,Giusto:2018ovt,Rastelli:2019gtj,Giusto:2019pxc,Goncalves:2019znr}, and beyond two-derivative supergravities \cite{Chester:2018lbz,Chester:2018dga,Chester:2018aca,Binder:2018yvd,Binder:2019jwn}  (see also \cite{Goncalves:2014ffa} for early developments). 

In this paper, we take a step further and study supergravity four-point functions of one-half BPS operators at one loop level. Thankfully, the technology developed in \cite{Alday:2016njk,Aharony:2016dwx}  allows us to readily  perform this task, which generalizes the generalized unitarity methods in flat space.\footnote{See also, {\it e.g.}, \cite{Giombi:2017hpr,Yuan:2017vgp,Yuan:2018qva,Bertan:2018afl,Carmi:2019ocp} for approaches which focus on individual loop diagrams.} A number of papers have appeared discussing supergravity one-loop correlators in $AdS_5\times S^5$ with small R-symmetry charges \cite{Alday:2017xua,Aprile:2017bgs,Aprile:2017xsp,Alday:2017vkk,Aprile:2017qoy,Aprile:2018efk}.\footnote{See also \cite{Alday:2018pdi,Chester:2019pvm} for recent progress on stringy corrections to loop amplitudes.} These results are impressive achievements because the expressions in position space are highly complicated. However, an important observation was made in \cite{Alday:2018kkw} that the $\mathbf{20'}$ operator four-point function continues to have simple analytic structures at one-loop level in Mellin space. The Mellin amplitude was found to consist of only simultaneous poles in the Mellin-Mandelstam variables with constant coefficients. This suggests that Mellin space perhaps is the most natural language to discuss these objects. In this paper, we will further demonstrate the simplicity of the one-loop correlators, by studying  in Mellin space an infinite family of correlators of the form $\langle \mathcal{O}^{SG}_2\mathcal{O}^{SG}_2\mathcal{O}^{SG}_p\mathcal{O}^{SG}_p\rangle$. Here $\mathcal{O}^{SG}_p$ stands for a one-half BPS operator with dimension $p$, and is dual to a scalar supergravity Kaluza-Klein mode with $S^5$ angular momentum $p$. The superscript ``{\it SG }'' refers to the fact that the field theory operators are in the {\it supergravity basis}, which we will elaborate on in Section \ref{secsugrabasis}. The reason for focusing on such correlators is that they enjoy certain properties which simplify the analysis. They allow us to explore in a simpler setting the analytic structures of the Mellin amplitudes. However, we believe that many of the analytic properties should also be valid for generic correlators. The main achievement of this paper is a systematic algorithm which computes the one-loop Mellin amplitudes of the $\langle \mathcal{O}^{SG}_2\mathcal{O}^{SG}_2\mathcal{O}^{SG}_p\mathcal{O}^{SG}_p\rangle$ correlators using the tree-level data. Moreover, we conjecture a closed form expression of the Mellin amplitudes for any $p$. We also study the flat space limit of these amplitudes, which gives perfect agreement with the flat space expectation. 
 
As we mentioned in the beginning, one motivation to study the holographic correlators is to discover unexpected structures, which may teach us new lessons about supersymmetric quantum gravity or strongly interacting CFTs. One curious emergent structure is the ten dimensional hidden conformal symmetry observed in the  $AdS_5\times S^5$ tree-level supergravity correlators \cite{Caron-Huot:2018kta}.\footnote{The same symmetry has also been observed in the leading order stringy corrections \cite{Drummond:2019odu}. A six dimensional version of the hidden symmetry has also been found in correlators from IIB supergravity on $AdS_3\times S^3\times K3$ \cite{Rastelli:2019gtj}. }  The authors of \cite{Caron-Huot:2018kta} showed that the result of \cite{Rastelli:2016nze,Rastelli:2017udc} can be resumed into a generating function which exhibits an $SO(10,2)$ symmetry. Moreover, the existence of the symmetry also explains the remarkable simplicity of the double-trace anomalous dimension formula \cite{Aprile:2018efk}, which comes from solving a complicated mixing problem. The fate of the hidden symmetry at one loop, however, is {\it a priori} unclear. While the hidden symmetry at tree level determines the leading logarithmic singularity at any loop order \cite{Caron-Huot:2018kta}, these singularities in general do not determine the full loop correlator even at one loop. Interestingly, we find nontrivial evidence that the one-loop correlators $\langle \mathcal{O}^{SG}_2\mathcal{O}^{SG}_2\mathcal{O}^{SG}_p\mathcal{O}^{SG}_p\rangle$ {\it are} completely determined by the hidden conformal symmetry. As we will define in a precise sense,  the full Mellin amplitudes are generated by the hidden symmetry, in terms of what we call the {\it pre-amplitude} which manifests the action of the symmetry (see Section \ref{subsecpreampli}). This gives rise to the hope that general correlators perhaps could also be fully determined using hidden symmetry. We will leave these speculations to future investigations. 

The rest of the paper is organized as follows. We start in Section \ref{generaldiscussion} with a general discussion of four-point functions at large $c$ and strong coupling. In Section \ref{secsugrabasis} we clarify the identification of the boundary operators with the one-half BPS supergravity states in the bulk by introducing the supergravity basis. In Section \ref{scfkinematics} we review the superconformal kinematics of the one-half BPS four-point functions. We review the implications of the ten dimensional hidden conformal symmetry in Section \ref{sechiddensymm}, and we introduce the Mellin representation in Section \ref{secMellinrep}. Section \ref{seconeloop} contains the main results for the one-loop correlators. We discuss the general structure of the $\langle \mathcal{O}_2^{SG}\mathcal{O}_2^{SG}\mathcal{O}_p^{SG}\mathcal{O}_p^{SG}\rangle$ one-loop correlators in Section \ref{subsec22pp}, and outline an algorithm for computing their Mellin amplitudes. We demonstrate this algorithm in Section \ref{subsecexamples}, by computing explicit examples with $p=2,3,4$. In Section \ref{subsecpreampli}, we expose the underlying structures of these Mellin amplitudes by introducing the pre-amplitudes. The action of the hidden symmetry is manifest on the pre-amplitudes, and this leads us to a conjecture for the $\langle \mathcal{O}_2^{SG}\mathcal{O}_2^{SG}\mathcal{O}_p^{SG}\mathcal{O}_p^{SG}\rangle$ one-loop Mellin amplitudes for any $p$. We comment on the generalization to higher-weight correlators in Section \ref{subsechigherweight}. In Section \ref{secflatspace} we study the flat space limit of the one-loop Mellin amplitudes.  We conclude in Section \ref{secconclusion} with a brief discussion of future directions. Additional technical details are relegated to the two appendices. 

{\bf Note:} While we were preparing this work for submission to the arXiv an independent work \cite{Aprile:2019rep} appeared, where a different algorithm in position space was developed. The authors demonstrated their algorithm by computing cases of correlators with low-lying conformal dimensions. Their explicit examples   complement our results for the one-parameter family $\langle \mathcal{O}_2^{SG}\mathcal{O}_2^{SG}\mathcal{O}_p^{SG}\mathcal{O}_p^{SG}\rangle$ in Mellin space.

\section{Four-Point Functions at Large $c$ and Strong Coupling}\label{generaldiscussion}
\subsection{The supergravity basis of one-half BPS operators}\label{secsugrabasis}
One-half BPS operators are super primaries of short representations of the superconformal group $PSU(2,2|4)$. These operators have zero Lorentz spin, Dynkin labels $[0,p,0]$ under the R-symmetry group $SU(4)$, and protected conformal dimensions $\Delta=p$. On the field theory side, we can explicitly write them down in terms of the six scalars $X^I$, $I=1,\ldots, 6$\;, of $\mathcal{N}=4$ SYM. A convenient way to enumerate them is to start with the {\it single-trace} one-half BPS operators 
\begin{equation}
O_p(x,t)=Tr X^{\{I_1}\ldots X^{I_p\}}t_{I_1}\ldots t_{I_p}\;, \quad  p=2,3,\ldots\;,
\end{equation}
and construct the $p${\it-trace} one-half BPS operators by taking normal-ordered products of $k$ single-trace operators and projecting to the symmetric traceless representation. Here we have used null vectors $t^I$ satisfying $t\cdot t=0$ to contract the R-symmetry indices, which makes the symmetric traceless property of the one-half BPS operators manifest. Note the distinction between single-trace and multi-trace is only sensible in the large $N$ limit -- at finite $N$ trace relations will give rise to relations among the operators. In this paper we will only focus on the limit where $N$ is large. The trace number however is not a useful notion on the dual AdS supergravity side, where it should be superseded by the {\it particle number} as we will elaborate below. 

To see this, let us first recall that supergravity fields from the Kaluza-Klein reduction are {\it not} mapped to single-trace operators under the duality dictionary \cite{Arutyunov:1999en,Arutyunov:2000ima,Rastelli:2017udc,Aprile:2018efk}. This is best illustrated by considering a simple example of a three-point function, namely $\langle O_2O_2O_4\rangle$. Three-point functions of one-half BPS operators are known to be independent of the 't Hooft coupling coupling \cite{Freedman:1998tz,Lee:1998bxa,Intriligator:1998ig,Intriligator:1999ff,Eden:1999gh,Petkou:1999fv,Howe:1999hz,Heslop:2001gp,Baggio:2012rr}. In the zero coupling limit, it can be computed from Wick contractions and yields a nonzero answer. At infinite 't Hooft coupling and tree level, three-point functions are computed as a three-point contact Witten diagram. However, we find that the effective Lagrangian of $AdS_5\times S^5$ IIB supergravity does not have a cubic vertex for three scalar fields with dimensions 2, 2 and 4. This seems to lead to a contradiction to the non-renormalization theorem. The resolution to the paradox \cite{Arutyunov:1999en,Arutyunov:2000ima} is that the Kaluza-Klein mode with dual conformal dimension 4 is actually a mixture of single-trace and double-trace operators (dubbed the ``extended CPO'' in \cite{Arutyunov:1999en,Arutyunov:2000ima})
\begin{equation}
\widetilde{O}_4(x,t)=O_4(x,t)+\mu :O_2O_2:(x,t)\;.
\end{equation} 
The mixing coefficient $\mu$ is fixed precisely by the condition that the three-point function $\langle \widetilde{O}_4 O_2 O_2\rangle$ is zero
\begin{equation}\label{zero422}
\langle \widetilde{O}_4(x_1,t_1) O_2 (x_2,t_2) O_2 (x_3,t_3)\rangle=0\;.
\end{equation}
The vanishing of $\langle \widetilde{O}_4 O_2 O_2\rangle$ is simple to understand as a consistency condition: the corresponding Witten diagram is divergent, so the coupling has to be zero in order for the effective action to be finite. However, should this property hold also at loop levels? To answer this, let us comprehend the physical meaning of the condition (\ref{zero422}). We note that the space of dimension 4 one-half BPS operators is two dimensional, and is spanned by $O_4$ and $:O_2O_2:$. The condition (\ref{zero422}) simply picks an orthogonal basis with new basis vectors $\widetilde{O}_4$ and $:O_2O_2:$
\begin{equation}\label{vanishextrm3pt}
\langle \widetilde{O}_4(x_1,t_1):O_2O_2:(x_2,t_2) \rangle=0\;.
\end{equation}
The operator $\widetilde{O}_4$ should be identified with a {\it single-particle} state in $AdS_5$, while $:O_2O_2:$ is identified with a {\it two-particle} state. The condition (\ref{vanishextrm3pt}) is the statement that the single-particle state and the two-particle state are orthogonal. Note that the orthogonality of operators with different particle numbers is a physical condition, and we should insist on this property at any perturbation order in the bulk. The condition (\ref{zero422}) therefore holds at any loop level.

The above discussion leads us to define a different basis for one-half BPS operators which is better suited for discussing the supergravity dual. We will refer to this basis as the {\it supergravity basis}. The construction of this basis goes as follows. We construct the single-particle operators as
\begin{equation}\label{defSGsingle}
O_p^{SG}=O_p+\sum_{a=2}^{\kappa}\sum_{\{q_i\}_{i=1}^a}\mu^{(a)}_{\{q_i\}}:O^{SG}_{q_1}\ldots O^{SG}_{q_a}:
\end{equation} 
where $\kappa=\left[\frac{p}{2}\right]\geq 2$ is the maximal number of traces $O_p^{SG}$ can accommodate. In the normal-ordered products, the representations have been projected to the symmetric traceless one to make them one-half BPS. This is conveniently implemented by using the same null vector for all operators. We sum over all the partitions $\{q_i\}$ satisfying the conditions
\begin{equation}
\mathbb{Z}\ni q_i\geq 2\;,\quad q_1\geq \ldots \geq q_a\;,\quad  \sum_i^a q_i=p\;.
\end{equation} 
The coefficients $\mu^{(a)}_{\{q_i\}}$ can be fixed by requiring that 
\begin{equation}\label{multiorthosing}
\langle :O^{SG}_{q_1}\ldots O^{SG}_{q_a}:(x_1,t_1) O_p^{SG}(x_2,t_2)\rangle=0\;,
\end{equation} 
for any $\{q_i\}$ satisfying the above partition conditions. These single-particle operators are identified with the Kaluza-Klein reduction of the supergravity fields. The above definition for single-particle operators is recursive.

 Let us unpack it by making a few comments. For any given $p$ in  (\ref{defSGsingle}), the normal-ordered products involve only single-particle operators with conformal dimensions smaller than $p$. Therefore we can start the construction from the lowest-lying operators and then increase the conformal dimensions. For $p=2,3$, the single-particle operators coincide with the single-trace operators
\begin{equation}
O_2^{SG}=O_2\;,\quad O_3^{SG}=O_3\;.
\end{equation} 
We start to encounter operator mixing at $p=4,5$
\begin{equation}
O_4^{SG}=O_4+\mu^{(2)}_{2,2}:O_2^{SG}O_2^{SG}:\;,\quad O_5^{SG}=O_5+\mu^{(2)}_{3,2}:O_3^{SG}O_2^{SG}:\;,
\end{equation}
and the mixing coefficients are solved by 
\begin{equation}
\langle :O_2^{SG}O_2^{SG}:O_4^{SG}\rangle=\langle :O_3^{SG}O_2^{SG}:O_5^{SG}\rangle=0\;.
\end{equation}
It is clear that for any value of $p$ the same number of conditions (\ref{multiorthosing}) is the same as the number of unknowns in (\ref{defSGsingle}), which ensures a solution. The condition (\ref{multiorthosing}) generalizes (\ref{vanishextrm3pt}) to a generic extremal correlation function, which is known to be true in tree-level supergravity. Note that extremal correlators are known to be protected \cite{DHoker:1999jke,Bianchi:1999ie,Eden:1999kw,Erdmenger:1999pz,Eden:2000gg}. The mixing coefficients $\mu^{(a)}_{\{q_i\}}$ are therefore coupling independent, and can be computed in the free theory.  Having defined the single-particle operators, we can define {\it multi-particle} operators by taking the normal-ordered products of the single-particle operators. The condition (\ref{multiorthosing}) is the statement that {\it multi-particle operators are orthogonal to single-particle operators}.

We should mention that the above notion of the supergravity basis is useful to any supergravity dual. In cases where the boundary field theory has a weakly-coupled limit, and the extremal correlators are protected by non-renormalization theorems, we can compute the mixing coefficients in the free theory limit. For example, we can similarly establish a relation between the supergravity states of IIB supergravity on $AdS_3\times S^3\times K3$ (or $T^4$), and the one-half BPS operators at the orbifold point. In the latter case operators are more naturally graded by the number of twist operators inserted, which is similar to the trace number in the $\mathcal{N}=4$ SYM case. 

For future convenience, we will also normalize the single-particle operators to have unit two-point function. We denote the normalized single-particle operators as $\mathcal{O}_p^{SG}$, and they satisfy
\begin{equation}
\langle\mathcal{O}_p^{SG}(x_1,t_1)\mathcal{O}_p^{SG}(x_2,t_2)\rangle=\left(\frac{t_{12}}{x_{12}^{2}}\right)^p
\end{equation}
where $t_{ij}=t_i\cdot t_j$, $x_{ij}=x_i-x_j$.

\subsection{Superconformal kinematics of four-point functions}\label{scfkinematics}
The focus of this paper is to study the four-point functions
\begin{equation}\label{hBPS4ptfun}
G_{p_1p_2p_3p_4}(x_i,t_i)\equiv\langle\mathcal{O}^{SG}_{p_1}(x_1,t_1)\mathcal{O}^{SG}_{p_2}(x_2,t_2)\mathcal{O}^{SG}_{p_3}(x_3,t_3)\mathcal{O}^{SG}_{p_4}(x_4,t_4)\rangle\;,
\end{equation}
 of one-half BPS operators which are single-particle operators in the supergravity basis. The objective of this section is to review the superconformal kinematics of these correlators.

\subsubsection{Solution to the superconformal constraints}
 We begin by considering covariance under R-symmetry. To ensure the correlator transforms covariantly under $SO(6)$ and with the correct R-symmetry charges, the null vectors can only appear in $G_{p_1p_2p_3p_4}$ as a linear combination of 
\begin{equation}
\prod_{i<j}t_{ij}^{\gamma_{ij}}\;,
\end{equation}
and the exponents are subject to the constraints
\begin{equation}
\gamma_{ij}\geq0\;,\quad \sum_{i\neq j}\gamma_{ij}=p_j\;.
\end{equation}
The monomials correspond to different Wick contractions in the free theory.  Superconformal symmetry places further constraints on the four-point functions, in the form of superconformal Ward identities \cite{Eden:2000bk,Nirschl:2004pa}. The solution to the superconformal Ward identities dictates that $G_{p_1p_2p_3p_4}$ can be written as the sum of a free part and a ``correction'' part
\begin{equation}\label{partnonren}
G_{p_1p_2p_3p_4}=G_{free,p_1p_2p_3p_4}+R\, H_{p_1p_2p_3p_4}\;,
\end{equation}
which is known in the literature as the partial non-renormalization theorem \cite{Eden:2000bk}. Here $G_{free,p_1p_2p_3p_4}$ is the four-point function in the free theory, and can be obtained by performing Wick contractions. The factor $R$ is crossing symmetric and is completely determined by superconformal symmetry to be
\begin{equation}
R=t_{12}^2t_{34}^2x_{13}^4x_{24}^4(1-z\alpha)(1-\bar{z}\alpha)(1-\bar{z}\alpha)(1-\bar{z}\bar{\alpha})\;.
\end{equation}
Here we have introduced the conformal cross ratios
\begin{equation}\label{defUV}
U=\frac{x_{12}^2x_{34}^2}{x_{13}^2x_{24}^2}=z\bar{z}\;,\quad V=\frac{x_{14}^2x_{23}^2}{x_{13}^2x_{24}^2}=(1-z)(1-\bar{z})\;,
\end{equation}
and analogously the R-symmetry cross ratios
\begin{equation}
\sigma=\frac{t_{13}t_{24}}{t_{12}t_{34}}=\alpha\bar{\alpha}\;,\quad \tau=\frac{t_{14}t_{23}}{t_{12}t_{34}}=(1-\alpha)(1-\bar{\alpha})\;.
\end{equation}
The function $H_{p_1p_2p_3p_4}$ is known as the {\it reduced correlator}, and encodes all the dynamical information. For (\ref{partnonren}) to be a nontrivial statement,  the reduced correlator must not contain any poles which cancel the zeros in $R$. Kinematically, we can view $H_{p_1p_2p_3p_4}$ as a four-point function with shifted conformal weights $p_i+2$ and R-symmetry weights $p_i-2$. It has, in particular, Bose symmetry (crossing symmetry in the case $p_i$ are identical) when the external operators are exchanged.

The above discussion relies only on superconformal symmetry, and applies to any one-half BPS four-point functions regardless of the mixing details among degenerate operators. However we should emphasize that $G_{p_1p_2p_3p_4}$ computed from supergravity must be identified with the four-point function of single-particle operators in the supergravity basis. The free correlator $G_{free,p_1p_2p_3p_4}$ in (\ref{partnonren}) gives different answers depending on whether the external one-half BPS operators are the single-trace operators $\mathcal{O}_p$ or the single-particle operators $\mathcal{O}_p^{SG}$.\footnote{\label{ftnoteambigu} The acute reader might wonder if the difference can be absorbed into a redefinition of $H_{p_1p_2p_3p_4}$, since the division of $G_{p_1p_2p_3p_4}$ into $G_{free,p_1p_2p_3p_4}$ and $H_{p_1p_2p_3p_4}$ is ambiguous. This however is not possible because we can set $\bar{\alpha}=1/\bar{z}$ (also known as the chiral algebra twist \cite{Beem:2013sza}) such that the contribution from the reduced correlator vanishes. The twisted correlator (with $\bar{\alpha}=1/\bar{z}$) is different for the two choices of external operators.} The mismatch of the free correlators was first observed in the explicit calculations from tree level supergravity \cite{Uruchurtu:2008kp,Uruchurtu:2011wh}, and is precisely accounted for by the change of operator basis. 

\subsubsection{Correlators as functions of cross ratios}
For later convenience, let us also extract some kinematic factors from the correlators such that we can write them as functions of cross ratios. We introduce the following convenient factor  
\begin{equation}
K_{p_1p_2p_3p_4}(x_i,t_i)=\left(\frac{t_{12}}{x_{12}^2}\right)^{\frac{p_1+p_2}{2}}\left(\frac{t_{34}}{x_{34}^2}\right)^{\frac{p_3+p_4}{2}}\left(\frac{x_{14}^2t_{24}}{x_{24}^2t_{14}}\right)^{\frac{p_2-p_1}{2}}\left(\frac{x_{14}^2t_{13}}{x_{13}^2t_{14}}\right)^{\frac{p_3-p_4}{2}}\;.
\end{equation}
We then define\footnote{Note the functions $\mathcal{G}$, $\mathcal{G}_{free}$, $\mathcal{H}$ defined here are slightly different from the ones introduced in \cite{Rastelli:2017udc} by some extra powers of $U$ and $V$.}
\begin{equation}
G_{p_1p_2p_3p_4}(x_i,t_i)=K_{p_1p_2p_3p_4}(x_i,t_i)\;\mathcal{G}_{p_1p_2p_3p_4}(U,V;\sigma,\tau)\;,
\end{equation}
\begin{equation}
G_{free,p_1p_2p_3p_4}(x_i,t_i)=K_{p_1p_2p_3p_4}(x_i,t_i)\;\mathcal{G}_{free,p_1p_2p_3p_4}(U,V;\sigma,\tau)\;,
\end{equation}
and
\begin{equation}
H_{p_1p_2p_3p_4}(x_i,t_i)=\left(t_{12}t_{34}x_{12}^2x_{34}^2\right)^{-2}K_{p_1p_2p_3p_4}(x_i,t_i)\;\mathcal{H}_{p_1p_2p_3p_4}(U,V;\sigma,\tau)\;.
\end{equation}
In terms of the functions of cross ratios, we can rewrite (\ref{partnonren}) as
\begin{equation}
\mathcal{G}_{p_1p_2p_3p_4}=\mathcal{G}_{free, p_1p_2p_3p_4}+\mathcal{R}U^{-2}\mathcal{H}_{p_1p_2p_3p_4}
\end{equation}
where 
\begin{equation}
\mathcal{R}=(1-z\alpha)(1-\bar{z}\alpha)(1-\bar{z}\alpha)(1-\bar{z}\bar{\alpha})\;.
\end{equation}

\subsubsection{Superconformal block decomposition}
As we mentioned in footnote \ref{ftnoteambigu}, the division between the protected part and the dynamic part is not without ambiguities. In fact, $\mathcal{G}_{p_1p_2p_3p_4}$ can be written in an alternative form which is more convenient for decomposing the correlator into superconformal blocks \cite{Dolan:2004iy,Bissi:2015qoa}
\begin{equation}\label{solscfWardid}
\begin{split}
\mathcal{G}_{p_1p_2p_3p_4}(z,\bar{z};\beta,\bar{\beta}){}&=k\, \chi (z,\alpha)\chi(\bar{z},\bar{\beta})+\frac{(z-\beta)(z-\bar{\beta})(\bar{z}-\beta)(\bar{z}-\bar{\beta})}{(\beta-\bar{\beta})(z-\bar{z})}\\
\times{}& \left(-\frac{\chi(\bar{z},\bar{\beta})f(z,\beta)}{\beta z(\bar{z}-\bar{\beta})}+\frac{\chi(\bar{z},\beta)f(z,\bar{\beta})}{\bar{\beta}z(\bar{z}-\beta)}+\frac{\chi(z,\bar{\beta})f(\bar{z},\beta)}{\beta\bar{z}(z-\bar{\beta})}-\frac{\chi(z,\beta)f(\bar{z},\bar{\beta})}{\bar{\beta}\bar{z}(z-\beta)}\right)\\
{}&+\frac{(z-\beta)(z-\beta)(\bar{z}-\beta)(\bar{z}-\bar{\beta})}{(z\bar{z})^2(\beta\bar{\beta})^2}\widetilde{\mathcal{H}}_{p_1p_2p_3p_4}(z,\bar{z};\beta,\bar{\beta})\;.
\end{split}
\end{equation} 
Here we have used a change of variables 
\begin{equation}
\beta=\frac{1}{\alpha}\;,\quad \bar{\beta}=\frac{1}{\bar{\alpha}}\;,
\end{equation}
and
\begin{equation}
k_{p_1p_2p_3p_4}=\mathcal{G}_{p_1p_2p_3p_4}(z,\bar{z};z,\bar{z})\;,
\end{equation}
\begin{equation}
f_{p_1p_2p_3p_4}(\bar{z},\bar{\beta})=\frac{\bar{\beta}\bar{z}}{\bar{z}-\bar{\beta}}\left(\mathcal{G}_{p_1p_2p_3p_4}(z,\bar{z};z,\bar{\beta})-k_{p_1p_2p_3p_4}\chi_{p_1p_2p_3p_4}(\bar{z},\bar{\beta})\right)\;,
\end{equation}
are coupling-independent functions, which are obtained by twisting the four-point correlator. The function $\chi_{p_1p_2p_3p_4}(z,\beta)$ is defined by
\begin{equation}
\chi_{p_1p_2p_3p_4}(z,\beta)=\left(\frac{z}{\beta}\right)^{\frac{\max\{|p_{21},|p_{34}||\}}{2}}\left(\frac{1-\beta}{1-z}\right)^{\frac{\max\{p_{21}+p_{34},0\}}{2}}
\end{equation}
where $p_{ij}=p_i-p_j$. The new form (\ref{solscfWardid}) partially reshuffles between $\mathcal{G}_{free,p_1p_2p_3p_4}$ and $\mathcal{H}_{p_1p_2p_3p_4}$. As a result, the superconformal block decomposition of the correlator now takes a simple form. However it is at the cost that $\widetilde{\mathcal{H}}_{p_1p_2p_3p_4}$ is no longer crossing symmetric. 

More precisely, we can write $f_{p_1p_2p_3p_4}$ and $\widetilde{\mathcal{H}}_{p_1p_2p_3p_4}$ as a sum over conformal blocks and R-symmetry blocks. The 4d conformal blocks are given by
\begin{equation}
g^{r,s}_{\Delta,\ell}(z,\bar{z})=\frac{z\bar{z}}{\bar{z}-z}\left[k^{r,s}_{\frac{\Delta-\ell-2}{2}}(z)k^{r,s}_{\frac{\Delta+\ell}{2}}(\bar{z})-k^{r,s}_{\frac{\Delta+\ell}{2}}(z)k^{r,s}_{\frac{\Delta-\ell-2}{2}}(\bar{z})\right]
\end{equation} 
where $r=p_{21}$, $s=p_{34}$, and 
\begin{equation}
k^{r,s}_h(z)=z^h {}_2F_1\left(h+\frac{r}{2},h+\frac{s}{2};2h,z\right)\;.
\end{equation}
The R-symmetry blocks are obtained by analytic continuations. We have 
\begin{equation}
Y_{m,n}^{r,s}(\beta,\bar{\beta})=(-1)^{m}g^{-r,-s}_{-n,m}(\beta,\bar{\beta})
\end{equation}
for the $SU(4)$ representation $[m,n-m,m]$, and 
\begin{equation}
y^{r,s}_m(\beta)=k^{-r,-s}_{-\frac{m}{2}}(\beta)\;,
\end{equation}
for the spin-$\frac{m}{2}$ representation of $SU(2)$. In terms of these building blocks, the functions  $\widetilde{\mathcal{H}}_{p_1p_2p_3p_4}$ and $f_{p_1p_2p_3p_4}$  are decomposed as 
\begin{equation}
\widetilde{\mathcal{H}}_{p_1p_2p_3p_4}=\sum_{\Delta,\ell,m,n}\mathcal{A}(\Delta,\ell,m,n)\, g^{p_{21},p_{34}}_{\Delta+4,\ell}(z,\bar{z})\,Y^{p_{21},p_{34}}_{m,n}(\beta,\bar{\beta})\;,
\end{equation}
and
\begin{equation}
f_{p_1p_2p_3p_4}(z,\beta)=\sum_{j=0}^\infty\sum_m \mu(j,m) k^{p_{21},p_{34}}_{1+\frac{m}{2}+j}y^{p_{21},p_{34}}_{\frac{m}{2}}(\beta)
\end{equation}
where $m$ is in the range $\max\{|p_{21}|,|p_{34}|\}\leq m\leq \min\{p_1+p_2,p_3+p_4\}-2$, and is even integers away from the lower bound. Long multiplets of $PSU(2,2|4)$ contribute only to $\widetilde{\mathcal{H}}$ but not to $f$. Short and semi-short multiplets, on the other hand, generally contribute to both $\widetilde{\mathcal{H}}$ and $f$. The precise expressions for contribution of each super multiplet can be found in \cite{Dolan:2004iy}.

Let us make a comment regarding the connection between the two equivalent forms (\ref{partnonren}) and (\ref{solscfWardid}) of writing the four-point function. When split according to (\ref{partnonren}), both $G_{free,p_1p_2p_3p_4}$ and $H_{p_1p_2p_3p_4}$ contribute to $\widetilde{\mathcal{H}}$. We respectively denote the contributions as $\widetilde{\mathcal{H}}_{free,p_1p_2p_3p_4}$ and $\widetilde{\mathcal{H}}_{dyn,p_1p_2p_3p_4}$. On the other hand, only $G_{free,p_1p_2p_3p_4}$ contributes to $f$. We can further split $\widetilde{\mathcal{H}}_{free,p_1p_2p_3p_4}$ into a short part $\widetilde{\mathcal{H}}^{short}_{free,p_1p_2p_3p_4}$ from the protected parts, and a long part $\widetilde{\mathcal{H}}^{long}_{free,p_1p_2p_3p_4}$ from long multiplets
\begin{equation}
\widetilde{\mathcal{H}}_{free,p_1p_2p_3p_4}=\widetilde{\mathcal{H}}^{short}_{free,p_1p_2p_3p_4}+\widetilde{\mathcal{H}}^{long}_{free,p_1p_2p_3p_4}\;.
\end{equation}
When speaking of the OPE coefficients of the long multiplets, we should take the total $\widetilde{\mathcal{H}}$, and with the contribution of short multiplets subtracted
\begin{equation}
\widetilde{\mathcal{H}}^{long}_{p_1p_2p_3p_4}=\widetilde{\mathcal{H}}_{dyn,p_1p_2p_3p_4}+\widetilde{\mathcal{H}}_{free,p_1p_2p_3p_4}-\widetilde{\mathcal{H}}^{short}_{free,p_1p_2p_3p_4}\;.
\end{equation}

\subsection{Implications of ten dimensional hidden conformal symmetry}\label{sechiddensymm}
The correlation function $G_{p_1p_2p_3p_4}$ admits an expansion in $1/c$ when the central charge $c=N^2-1$ is large. It is convenient to define an expansion parameter
\begin{equation}\label{defa}
a=\frac{1}{4c}=\frac{1}{N^2-1}\;,
\end{equation}
and we can write the expansion as 
\begin{equation}
G_{p_1p_2p_3p_4}=G^{(0)}_{p_1p_2p_3p_4}+a\,G^{(1)}_{p_1p_2p_3p_4}+a^2\,G^{(2)}_{p_1p_2p_3p_4}+\ldots
\end{equation}
Thanks to (\ref{partnonren}), computing $G_{p_1p_2p_3p_4}$ reduces to the task of calculating $H_{p_1p_2p_3p_4}$, which admits a similar expansion
\begin{equation}
H_{p_1p_2p_3p_4}=a\,H^{(1)}_{p_1p_2p_3p_4}+a^2\,H^{(2)}_{p_1p_2p_3p_4}+\ldots\;.
\end{equation}
Note that $H^{(0)}_{p_1p_2p_3p_4}$ is missing because the leading order correlator $G^{(0)}_{p_1p_2p_3p_4}$ coincides with $G^{(0)}_{free,p_1p_2p_3p_4}$ due to large $N$ factorization. Using the AdS/CFT correspondence, obtaining $H^{(L+1)}_{p_1p_2p_3p_4}$ requires to compute $AdS_5\times S^5$ supergravity to $L$ loops. For $L=0$, this corresponds to tree level supergravity, and all one-half BPS four-point functions $H^{(1)}_{p_1p_2p_3p_4}$ were obtained by solving an algebraic bootstrap equation \cite{Rastelli:2016nze,Rastelli:2017udc}. In \cite{Caron-Huot:2018kta} it was realized that the results of \cite{Rastelli:2016nze,Rastelli:2017udc} can be elegantly organized in terms of a hidden ten dimensional conformal symmetry. In this section, we briefly recall and comment on the consequence of this hidden structure.

The hidden symmetry is manifested at tree level by a generating function which is obtained by lifting the $p_i=2$ correlator into ten dimensions. More precisely, we start from the reduced correlator $H^{(1)}_{2222}$ which is a function of $x_{ij}^2$ only. We define the generating function $\mathbf{H}$ by replacing the argument $x_{ij}^2$ with $x_{ij}^2+t_{ij}$
\begin{equation}\label{treegen}
\mathbf{H}(x_i,t_i)=H^{(1)}_{2222}(x_{ij}^2+t_{ij})\;.
\end{equation}
The higher-weight reduced correlators $H^{(1)}_{p_1p_2p_3p_4}$ are obtained by Taylor expanding $\mathbf{H}(x_i,t_i)$ in $t_{ij}$ and collect all the monomials allowed in $H^{(1)}_{p_1p_2p_3p_4}$. When rewritten in terms of the cross ratios, one obtains differential relations
\begin{equation}\label{treediffrelation}
\mathcal{H}^{(1)}_{p_1p_2p_3p_4}=\mathcal{D}_{p_1p_2p_3p_4}\mathcal{H}^{(1)}_{2222}\;.
\end{equation}
It is easy to check that this way of generating the differential operators $\mathcal{D}_{p_1p_2p_3p_4}$ is equivalent to the original prescription in \cite{Caron-Huot:2018kta} via contour integrals. 

The ten dimensional hidden conformal symmetry also reveals interesting structures in the theory spectrum at infinite coupling. These structures of the spectrum have implications which extend well beyond the tree level. In the supergravity approximation, all unprotected operators are formed as non-BPS multi-particle states, and are degenerate in the strict $c\to\infty$ limit. The degeneracy of the two-particle spectrum is lifted at the next order by the averaged anomalous dimensions which are encoded in the $\log U$ coefficient of $\mathcal{H}^{(1)}_{p_1p_2p_3p_4}$. The anomalous dimensions can be systematically extracted from $\mathcal{G}^{(0)}_{p_1p_2p_3p_4}$ and $\mathcal{H}^{(1)}_{p_1p_2p_3p_4}$ after diagonalizing the mixing matrix \cite{Aprile:2017xsp,Aprile:2017qoy}. Suprisingly, solving the mixing problem becomes automatic if one decomposes the reduced correlator in terms of ten dimensional conformal blocks, as was demonstrated in \cite{Caron-Huot:2018kta}. This interesting feature not only provides a more efficient way to obtain the anomalous dimensions, but also gives a prediction about the leading logarithmic singularity of the reduced correlator. Because the use of ten dimensional conformal blocks diagonalizes the spectrum, it follows that the $\log^q U$ coefficient of  $\mathcal{H}^{(q)}_{p_1p_2p_3p_4}$,  related to the $q$-th power of the tree level anomalous dimension $\gamma^{(1)}$,  is given by \cite{Caron-Huot:2018kta}
\begin{equation}\label{lls}
\mathcal{H}^{(q)}_{p_1p_2p_3p_4}\big|_{\log^q U}=\left[\Delta^{(8)}\right]^{q-1}\cdot \mathcal{D}_{p_1p_2p_3p_4}\cdot r^{(q)}(z,\bar{z})\;.
\end{equation}
Here $\Delta^{(8)}$ is an eighth-order differential operator of which we record the explicit form in Appendix \ref{app:diffops}. The operators $\mathcal{D}_{p_1p_2p_3p_4}$ are the same ones that appeared in (\ref{treediffrelation}). Finally, $r^{(q)}(z,\bar{z})$ are the seed data which do not depend on the external weights $\{p_i\}$ but only on the loop level. Explicit expressions for $r^{(q)}(z,\bar{z})$ can be found in Appendix C of \cite{Caron-Huot:2018kta}.\footnote{Our notation deviates from \cite{Caron-Huot:2018kta}, where $r^{(k)}(z,\bar{z})$ is denoted as $\mathcal{D}_{(3)}\cdot h^{(k)}(z)$.}

We will focus on $q=2$, {\it i.e.}, one loop level, in the rest of the paper. It will turn out to be useful to view $r^{(2)}(z,\bar{z})$ as the $\log^2 U$ coefficient of a correlator-like object, to wit
\begin{equation}
r^{(2)}(z,\bar{z})=\lambda_{2222}\big|_{\log^2U}\;,
\end{equation} 
or equivalently, 
\begin{equation}
\lambda_{2222}=r^{(2)}(z,\bar{z})\log^2U+\ldots
\end{equation} 
where $\ldots$ contains subleading logarithmic singularities in $U$. Similarly, we can introduce the higher-weight analogue of $\lambda_{2222}$ via the identification
\begin{equation}
\mathcal{D}_{p_1p_2p_3p_4}\cdot r^{(2)}(z,\bar{z})=\lambda_{p_1p_2p_3p_4}\big|_{\log^2U}\;.
\end{equation}
Note that $\mathcal{D}_{p_1p_2p_3p_4}$ cannot act on $\log^2 U$ because it has to preserve the leading logarithmic singularity. The relations between the $\log^2 U$ coefficients suggests that the correlators $\lambda_{p_1p_2p_3p_4}$ and $\lambda_{2222}$ should be related by the hidden ten dimensional conformal symmetry in the same fashion (\ref{treegen}) as at tree level. 

\subsection{Mellin representation}\label{secMellinrep}
The Mellin representation formalism \cite{Mack:2009mi,Penedones:2010ue,Fitzpatrick:2011ia} is an efficient language for discussing holographic correlators, and manifests the analytic structures. We will follow \cite{Rastelli:2016nze,Rastelli:2017udc} where the Mandelstam-Mellin variables simply permute under the action of Bose symmetry. 

We define the Mellin representation of the reduced correlator to be\footnote{Here we adopted the notation from \cite{Rastelli:2016nze,Rastelli:2017udc} and denoted the reduced amplitude with a tilde. This notation was used to distinguish it from the {\it full} Mellin amplitude $\mathcal{M}_{p_1p_2p_3p_4}$ that appears in the inverse Mellin transformation of the full correlator $\mathcal{G}_{p_1p_2p_3p_4}$. However, the latter amplitude will make no appearance in this paper.} 
\begin{equation}\label{calHMellin}
\begin{split}
\mathcal{H}_{p_1p_2p_3p_4}={}&\int \frac{dsdt}{(4\pi i)^2}U^{\frac{s+4}{2}}V^{\frac{t-p_2-p_3}{2}}\widetilde{\mathcal{M}}_{p_1p_2p_3p_4}(s,t;\sigma,\tau)\Gamma[\tfrac{p_1+p_2-s}{2}]\Gamma[\tfrac{p_3+p_4-s}{2}]\\
{}&\times \Gamma[\tfrac{p_1+p_4-t}{2}]\Gamma[\tfrac{p_2+p_3-t}{2}]\Gamma[\tfrac{p_1+p_3-u}{2}]\Gamma[\tfrac{p_2+p_4-u}{2}]\;.
\end{split}
\end{equation}
Here $s+t+u=p_1+p_2+p_3+p_4-4$.\footnote{We have slightly changed the notation compared to \cite{Rastelli:2017udc}. In \cite{Rastelli:2017udc}, the Mandelstam variables $s$, $t$, $u$ satisfying $s+t+u=p_1+p_2+p_3+p_4$ were defined for $\mathcal{G}_{p_1p_2p_3p_4}$, and they permute under crossing. A shifted variable $\tilde{u}=u-4$ was defined for $\mathcal{H}_{p_1p_2p_3p_4}$ such that $s$, $t$, $\tilde{u}$ are permuted. Since this paper we will mostly talk about $\mathcal{H}_{p_1p_2p_3p_4}$ at one loop, we drop the tilde on $u$ in order to avoid overloading the notations.} Using this formalism, the tree level reduced correlators admit an extremely simple representation. The Mellin amplitudes take the form \cite{Rastelli:2016nze,Rastelli:2017udc}
\begin{equation}\label{treeMellin}
\widetilde{\mathcal{M}}^{(1)}_{p_1p_2p_3p_4}=\sum_{i,j,k}\frac{a_{ijk}\sigma^i\tau^j}{(s-s_M+2k)(t-t_M+2j)(u-u_M+2i)}
\end{equation}
where the summation is over all the R-symmetry structures. The range of the summation is constrained by the condition $i+j+k=\min\{p_{\min},\frac{p_1+p_2+p_3+p_4-2p_{\max}}{2}\}-2$\;, where $p_{\min}$ and $p_{\max}$ are respectively the smallest and largest weights of $p_i$.
The values of $s_M$, $t_M$, $u_M$ are given by
\begin{equation}
s_M=\min\{p_1+p_2,p_3+p_4\}-2\;,\;\; t_M=\min\{p_1+p_4,p_2+p_3\}-2\;,\;\; u_M=\min\{p_1+p_3,p_2+p_4\}-2\;,
\end{equation}
which ensure the poles in (\ref{treeMellin}) are in the region bounded by the poles from the Gamma functions.

Let us also define the Mellin representation for $\lambda_{p_1p_2p_3p_4}$. Since the differential operator $\Delta^{(8)}$ can be written in terms of Casimir operators, it preserves the conformal and R-symmetry weights. We therefore  have the same representation for $\lambda_{p_1p_2p_3p_4}$
\begin{equation}\label{lambdaMellin}
\begin{split}
\lambda_{p_1p_2p_3p_4}={}&\int \frac{dsdt}{(4\pi i)^2}U^{\frac{s+4}{2}}V^{\frac{t-p_2-p_3}{2}}\mathcal{L}_{p_1p_2p_3p_4}(s,t;\sigma,\tau)\Gamma[\tfrac{p_1+p_2-s}{2}]\Gamma[\tfrac{p_3+p_4-s}{2}]\\
{}&\times \Gamma[\tfrac{p_1+p_4-t}{2}]\Gamma[\tfrac{p_2+p_3-t}{2}]\Gamma[\tfrac{p_1+p_3-u}{2}]\Gamma[\tfrac{p_2+p_4-u}{2}]\;.
\end{split}
\end{equation}

\section{Supergravity One-Loop Correlators in Mellin Space}\label{seconeloop}
In this section we demonstrate the simplicity of $AdS_5\times S^5$ supergravity Mellin amplitudes at one loop. We will mostly focus on the class of correlators with $p_1=p_2=2$, $p_3=p_4=p$, for which a number of simplifications occur and a general solution is easier to obtain. In Section \ref{subsec22pp} we discuss the structure of their Mellin amplitudes. We describe an algorithm which generalizes the approach of \cite{Alday:2018kkw} and computes Mellin amplitudes using implications of the hidden ten dimensional conformal symmetry and other tree-level data. This algorithm is implemented in Section \ref{subsecexamples} for several explicit examples. In Section \ref{subsecpreampli}, we introduce the pre-amplitudes, which are related to the Mellin amplitudes via differential operators. We find a general formula for the pre-amplitudes, and it gives the one-loop Mellin amplitude with arbitrary $p$. When applied to more general correlators, our algorithm captures only part of answer because new structures will emerge. We will comment about the general cases in Section \ref{subsechigherweight}, and discuss what extra information is needed to find the complete answer.

\subsection{General structure of $\langle \mathcal{O}_2^{SG}\mathcal{O}_2^{SG}\mathcal{O}_p^{SG}\mathcal{O}_p^{SG}\rangle$  and outline of the algorithm}\label{subsec22pp}
\subsubsection*{Simplifications of $\langle \mathcal{O}_2^{SG}\mathcal{O}_2^{SG}\mathcal{O}_p^{SG}\mathcal{O}_p^{SG}\rangle$}
The four-point functions $\langle \mathcal{O}_2^{SG}\mathcal{O}_2^{SG}\mathcal{O}_p^{SG}\mathcal{O}_p^{SG}\rangle$ are the ideal starting point for exploring non-planar correlators. By focusing on these correlators, we can avoid some complications which appear in the general Kaluza-Klein correlators, while still learn about the essential physics of $AdS_5\times S^5$ supergravity at one loop. The simplification of $\langle \mathcal{O}_2^{SG}\mathcal{O}_2^{SG}\mathcal{O}_p^{SG}\mathcal{O}_p^{SG}\rangle$ correlators are: 
\begin{itemize}
\item The reduced correlator $\mathcal{H}_{22pp}$ has no R-symmetry dependence. This is clear from the discussion of superconformal kinematics in Section \ref{scfkinematics}.
\item Pairwise identical operators give rise to crossing symmetry between the t and u channel.  
\item The free correlator $\mathcal{G}_{free,22pp}$ is $1/c$-exact. We will prove this fact in Appendix \ref{apponeovercexact}.
\end{itemize}
While the first two points lead to obvious simplifications, the consequence of the last bullet point deserves some comments. 
It follows from the $1/c$-exactness of the free correlator (recall $a$ is related to $c$ via (\ref{defa}))
\begin{equation}
\mathcal{G}_{free,22pp}=\mathcal{G}^{(0)}_{free,22pp}+a\,\mathcal{G}^{(1)}_{free,22pp}\;,
\end{equation}  
that for $q\geq 2$
\begin{equation}
\mathcal{H}^{(q)}_{free,22pp}=0\;,
\end{equation}
This implies that 
\begin{equation}
\widetilde{\mathcal{H}}^{(q)}_{22pp}=\widetilde{\mathcal{H}}^{(q),long}_{22pp}=\mathcal{H}^{(q)}_{22pp}\;,\quad q\geq 2
\end{equation}
where we recall that $\widetilde{\mathcal{H}}^{(q)}_{22pp}$ was defined in the solution to the superconformal Ward identity  (\ref{solscfWardid}), which admits a convenient decomposition into superconformal blocks; and $\widetilde{\mathcal{H}}^{(q),long}_{22pp}$ is the contribution to $\widetilde{\mathcal{H}}^{(q)}_{22pp}$ from all long multiplets. Similarly, for the other two channels we also have 
\begin{equation}
\widetilde{\mathcal{H}}^{(q)}_{2pp2}=\widetilde{\mathcal{H}}^{(q),long}_{2pp2}=\mathcal{H}^{(q)}_{2pp2}\;,\quad q\geq 2\;,
\end{equation}
\begin{equation}
\widetilde{\mathcal{H}}^{(q)}_{2p2p}=\widetilde{\mathcal{H}}^{(q),long}_{2p2p}=\mathcal{H}^{(q)}_{2p2p}\;,\quad q\geq 2\;.
\end{equation}
Should $\mathcal{G}^{(q)}_{free,22pp}$ have not vanished, the free theory contribution $\widetilde{\mathcal{H}}^{(q)}_{free,22pp}$ would be nonzero. It may in particular  contain contributions to $\widetilde{\mathcal{H}}^{(2),long}_{22pp}$ from operators with twist $\tau=2$. Such operators however are unphysical in the supergravity limit, where long operators are multi-particle operators and have a minimal engineering twist 4. The presence of such contribution in $\widetilde{\mathcal{H}}^{(q)}_{free,22pp}$ would require $\mathcal{H}^{(q)}_{22pp}$ to have exactly opposite contributions such that the spurious long operators do not appear in the sum.\footnote{Such cancellation can be used to fix the overall normalization of $\mathcal{H}^{(q)}_{p_1p_2p_3p_4}$, as was done at the $q=1$ tree level in \cite{Dolan:2006ec,Aprile:2017xsp,Aprile:2017qoy,Aprile:2018efk}.} Thanks to the fact that $\mathcal{G}^{(q)}_{free,22pp}=0$ for $q\geq 2$, we see that $\mathcal{H}^{(q)}_{22pp}$ contains no $\tau=2$ operators. Similarly,  the vanishing $\mathcal{G}^{(q)}_{free,22pp}$ implies $\widetilde{\mathcal{H}}^{(q),long}_{free,2pp2}$ and $\widetilde{\mathcal{H}}^{(q),long}_{free,2p2p}$ are zero, and in particular the absence of long operator contributions with twist $\tau<p+2$. Note that such operators are also spurious for these supergravity correlators: for $p_1+p_2<p+2$, correlators $\langle\mathcal{O}^{SG}_2\mathcal{O}^{SG}_p\mathcal{O}^{SG}_{p_1}\mathcal{O}^{SG}_{p_2}\rangle$ have to vanish because of R-symmetry violation, or for being extremal or next-to-extremal; this implies that no double-particle long operators with $\tau<p+2$ can appear in the $\mathcal{O}^{SG}_2\times\mathcal{O}^{SG}_p$ OPE in the supergravity limit. All in all, we conclude that $\mathcal{H}^{(q)}_{22pp}=0$ contains only physical long operators for $q\geq 2$.   

\subsubsection*{Structure of the one-loop Mellin amplitude}
The above consideration immediately sets a lower bound for the pole ranges in the Mellin integrand. In Mellin space exchanged operators are manifested as poles in the integrand of the inverse Mellin transformation. These poles come in infinite series with even integer spacing, and the position of the leading pole coincides with the twist of the exchanged operator. Because $\mathcal{H}^{(2)}_{22pp}$ contains only physical long operators which are double-particle states, the poles can only be selected from the following set
\begin{equation}\label{lowerbounds}
s=4,6,\ldots\;,\quad t,u=p+2,p+4,\ldots\;.
\end{equation}
Moreover, the integrand of the inverse Mellin transformation cannot have more than triple poles. This is because the one-loop reduced correlator can have at most $\log^2U$ logarithmic singularity in the small $U$ expansion, which is associated with the squared tree-level anomalous dimension $\left(\gamma^{(1)}\right)^2$ from the $1/c$ expansion.

\begin{figure}[htbp]
\begin{center}
\includegraphics[width=0.8\textwidth]{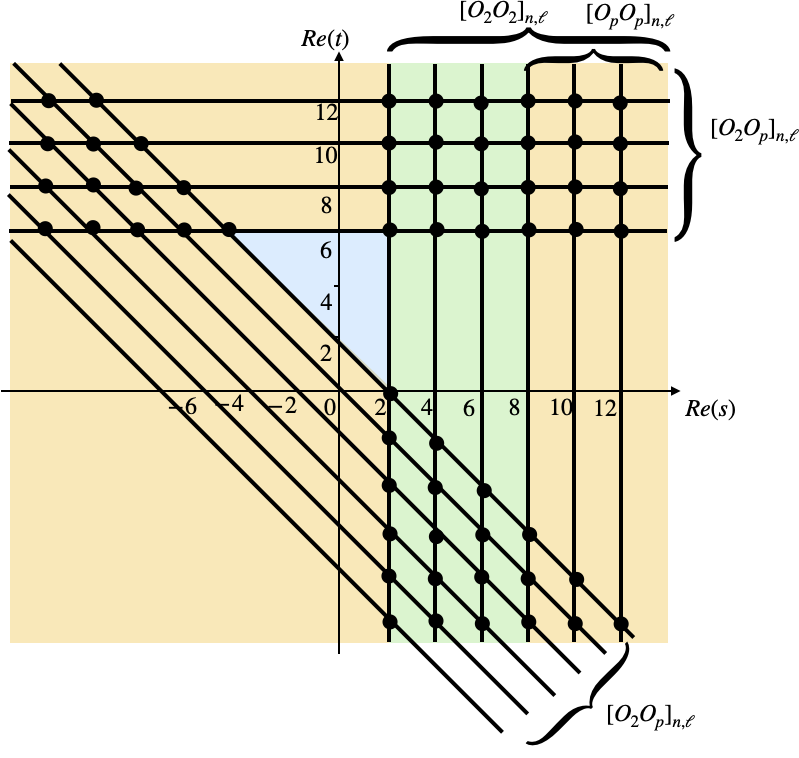}
\caption{Pole structures of the Mellin amplitude $\widetilde{\mathcal{M}}^{(2)}_{22pp}$ (for the example of $p=4$). The poles of the Gamma functions are represented by the straight lines, and the simultaneous poles in the ansatz (\ref{ansatz22pp}) are denoted by the dots at the intersections of the lines. The light blue region is where the poles of the tree level supergravity solution (\ref{treeMellin}) sit, which is bounded by the Gamma function poles. We have used the green color for the region $2\leq Re(s)< 8$ to emphasize that the two series of Gamma function poles do not overlap in this region, which corresponds to the long operators $[\mathcal{O}_2\mathcal{O}_2]_{n,\ell}$ with $n=2,\ldots p-1$.}
\label{polestructure}
\end{center}
\end{figure}

Further insight on the structure of Mellin amplitudes can be gained from the $p=2$ case computed in \cite{Alday:2018kkw}. It is observed that the Mellin amplitude $\widetilde{\mathcal{M}}_{2222}$ consists only of simultaneous simple poles with constant residues. The location of the poles in the Mellin amplitude overlap completely with the poles in the Gamma function factors, and correspond to all the double-particle long operators. We are therefore motivated to propose the following Mellin amplitudes ansatz for $\mathcal{H}^{(2)}_{22pp}$
\begin{equation}\label{ansatz22pp}
\begin{split}
\widetilde{\mathcal{M}}^{(2)}_{22pp}={}&\sum_{m,n=0}^\infty\frac{c_{mn}^u}{(s-4-2m)(t-(2+p)-2n)}+\sum_{m,n=0}^\infty\frac{c_{mn}^t}{(s-4-2m)(u-(2+p)-2n)}\\
{}&+\sum_{m,n=0}^\infty \frac{c_{mn}^s}{(t-(2+p)-2n)(u-(2+p)-2m)}
\end{split}
\end{equation}
where $c^t_{mn}=c^u_{mn}$ by crossing symmetry.\footnote{Readers familiar with \cite{Alday:2018kkw} might recall that the Mellin amplitude with $p=2$ (and $p>2$ as we will see) is divergent in the sum over $m$, $n$. Strictly speaking, the sum requires a regularization as explained in \cite{Alday:2018kkw} and also in Section \ref{secflatspace}, and differently regularized amplitudes differ by a contact term. What we compute in this section is the residues of the finite part, which is extracted from the small $U$, $V$ expansion of the correlator.  This calculation is independent of regularization. Note that the small $U$, $V$ also improve the convergence. See Section 3.4 of \cite{Alday:2018kkw} for details.}  Note that in contrast to $p=2$, the case with general $p$ contains two separate Gamma functions in the s-channel, {\it i.e.}, $\Gamma[\frac{4-s}{2}]$ and $\Gamma[\frac{2p-s}{2}]$. The Gamma functions have a finite range in which the poles do not overlap. This non-overlapping range corresponds to the double-particle long operators $[\mathcal{O}_2\mathcal{O}_2]_{n,\ell}$ which only become degenerate with $[\mathcal{O}_p\mathcal{O}_p]_{n',\ell}$ when $n\geq p-2$. We should emphasize that in the ansatz (\ref{ansatz22pp}) the poles in $s$ overlap with all the poles of $\Gamma[\frac{4-s}{2}]$, which start at the lower bound in (\ref{lowerbounds}). The pole structure of this ansatz is illustrated by Figure \ref{polestructure}.

We will now spend the rest of this section to outline a strategy to systematically solve the ansatz. As we will show, the ansatz at one loop is intimately related to the data that one can extract from the tree level. We will explain how these relations will enable us to use the tree level data to confirm that (\ref{ansatz22pp}) is the correct answer.

Solving the ansatz amounts to finding all the coefficients $c^s_{mn}$, $c^t_{mn}$, $c^u_{mn}$. These coefficients can be fixed from the leading singularity of the reduced correlator, which can be computed by using hidden symmetry relation (\ref{lls}). More precisely, the coefficients are encoded in the $\log^2U\log^2V$ (or leading $\log^2U\log V$) coefficients of the reduced correlator. For example, in 
\begin{equation}
\mathcal{H}^{(2)}_{22pp}=\int_{-i\infty}^{i\infty}\frac{dsdt}{(4\pi i)^2}U^{\frac{s+4}{2}}V^{\frac{t-p-2}{2}}\widetilde{\mathcal{M}}^{(2)}_{22pp}(s,t)\Gamma[\tfrac{4-s}{2}]\Gamma[\tfrac{2p-s}{2}]\Gamma[\tfrac{2+p-t}{2}]^2\Gamma[\tfrac{2+p-u}{2}]\;,
\end{equation}
the simultaneous poles in the first sum of (\ref{ansatz22pp}), with $m\geq p-2$ and $n\geq 0$, give rise to triple poles in $s$ and $t$ for the inverse Mellin integrand. Upon closing the $s$ and $t$ contours to the right and taking the residues, the simultaneous triple poles generate $\log^2U\log^2V$ singularities (to be precise, each pair of simultaneous pole also gets a $U^mV^n$ factor from the $U^{\frac{s+4}{2}}V^{\frac{t-p-2}{2}}$ factor in the integrand). By comparing with the $\log^2V$ coefficient of $\mathcal{H}^{(2)}_{22pp}\big|_{\log^2U}$, we can uniquely fix $c^u_{mn}$ for $m\geq p-2$. Similarly, we can consider
\begin{equation}
\mathcal{H}^{(2)}_{2pp2}=\int_{-i\infty}^{i\infty}\frac{dsdt}{(4\pi i)^2}U^{\frac{s+4}{2}}V^{\frac{t-2p}{2}}\widetilde{\mathcal{M}}^{(2)}_{2pp2}(s,t)\Gamma[\tfrac{4-t}{2}]\Gamma[\tfrac{2p-t}{2}]\Gamma[\tfrac{2+p-s}{2}]^2\Gamma[\tfrac{2+p-u}{2}]
\end{equation}
where we have used crossing symmetry to map $\widetilde{\mathcal{M}}^{(2)}_{22pp}$ into $\widetilde{\mathcal{M}}^{(2)}_{2pp2}$. The $\log^2U\log^2V$ coefficient now receives contributions only from the second sum in (\ref{ansatz22pp}), and it has been mapped into
\begin{equation}
\sum_{m,n=0}^\infty\frac{c_{mn}^t}{(t-4-2m)(s-(2+p)-2n)}\;.
\end{equation}  
For $n\geq 0$, and $m\geq p-2$, the integrand has triple poles and gives $\log^2U\log^2V$  singularities. For $0\leq m< p-2$, however, the integrand has only double poles in $t$ and generates at most $\log^2U\log V$. Moreover, these $\log^2U\log V$ terms are multiplied with the singular powers $V^{a-p}$ with $a=2,3,\ldots p-1$ because of the $V^{\frac{t-2p}{2}}$ factor in the integrand. This allows us to isolate them from the subleading $\log^2U\log V$ singularities which arise from the poles with $n\geq 0$, and $m\geq p-2$. Therefore we can  determine the remaining $c_{mn}^t$ with $m=0,1,\ldots p-1$, by comparing the residues with the leading $\log V$ coefficients in $\mathcal{H}^{(2)}_{2pp2}\big|_{\log^2U}$ in the small $V$ expansion. By crossing symmetry, we have $c_{mn}^u=c_{mn}^t$, and this fully determines $c_{mn}^u$. Finally, by comparing the $\log^2U\log^2V$ coefficient from $\widetilde{\mathcal{M}}^{(2)}_{2p2p}$ with the $\log^2V$ coefficient of $\mathcal{H}^{(2)}_{2p2p}\big|_{\log^2U}$, we can solve the coefficients $c^s_{mn}$.

One may wonder if the above ansatz with only simultaneous poles is sufficient. Moreover, one can ask if single poles of the form 
\begin{equation}\label{singlepole}
\sum_{m=0}^{p-1}\frac{f_m(t)}{s-2m}\;,
\end{equation}
with regular $f_m(t)$ should be added. We will answer this question in two steps. We first insert the above coefficients into the Mellin ansatz, and integrate to reproduce the {\it full} $V$-dependence in $\mathcal{H}^{(2)}_{22pp}\big|_{\log^2U}$, $\mathcal{H}^{(2)}_{2pp2}\big|_{\log^2U}$ and $\mathcal{H}^{(2)}_{2p2p}\big|_{\log^2U}$. Note this includes not only the $\log^2V$ and leading $\log V$ terms which had been used to fix the coefficients, but also all the other terms with and without $\log V$ singularity. Reproducing the full leading logarithmic singularities in all three channels is highly nontrivial -- it guarantees that (\ref{ansatz22pp}) is correct in the region with $s\geq 2p$ and $t,u\geq 2+p$. This still allows the possibility of adding single poles (\ref{singlepole}) in the region $4\leq s< 2p$, where only double poles in $s$ are present in the inverse Mellin integrand (and only double poles in $t$, so at most $\log U\log V$). To see whether we can exclude these additional terms, we take the $s$-residues in this region. If the ansatz is complete, the leading $\log U$ coefficient should have the CFT interpretation of
\begin{equation}\label{treeSGlogU}
\sum_{\tau=4}^{2p-2}\tfrac{1}{2}\sum_{i} C^{(0)}_{22K_{\tau,\ell,i}}\gamma^{(1)}_{\tau,\ell,i}\, C^{(1)}_{ppK_{\tau,\ell,i}} g_{\tau+\ell+4,\ell}(z,\bar{z})+\mathcal{O}(U^{p+2})
\end{equation}
where $K_{\tau,\ell,i}$ denotes the long operator with twist $\tau$, spin $\ell$, degeneracy label $i$, and tree level anomalous dimension $\gamma^{(1)}_{\tau,\ell,i}$. These long operators appear in the OPE of $\mathcal{O}_2^{SG}$ with $\mathcal{O}_2^{SG}$ at disconnected level with OPE coefficients $C^{(0)}_{22K_{\tau,\ell,i}}$; but only appear in the OPE of $\mathcal{O}_p^{SG}$ with $\mathcal{O}_p^{SG}$ with sub-leading order OPE coefficients $C^{(1)}_{ppK_{\tau,\ell,i}}$, which can be computed from tree level supergravity correlators. 
The zeroth order OPE coefficient $C^{(0)}_{22K_{\tau,\ell,i}}$, or more generally, $C^{(0)}_{rrK_{\tau,\ell,i}}$ for $r\leq \frac{\tau}{2}$ can be obtained from solving the mixing problem that gives the tree level anomalous dimensions \cite{Aprile:2017xsp,Aprile:2017qoy}. The coefficients $C^{(1)}_{ppK_{\tau,\ell,i}}$ for $p\geq \frac{\tau}{2}$, on the other hand, can be obtained by from the $U^{\frac{\tau}{2}+2}$, $\tau=4,\ldots 2p-2$ coefficients in the small $U$ expansion of the tree level correlators $\mathcal{H}^{(1)}_{rrpp}$ with $2\leq r\leq p-1$. These expansion coefficients have the interpretation of 
\begin{equation}
\sum_{\tau=2r}^{2p-2}\sum_{i} C^{(0)}_{rrK_{\tau,\ell,i}} C^{(1)}_{ppK_{\tau,\ell,i}}g_{\tau+\ell+4,\ell}(z,\bar{z})+\mathcal{O}(U^{p+2})
\;,
\end{equation}
from which we can solve $C^{(1)}_{ppK_{\tau,\ell,i}}$ with the input of $C^{(0)}_{rrK_{\tau,\ell,i}}$ from the solution of the mixing problem. With all the OPE coefficients and anomalous dimensions at hand, we can straightforwardly resum the $\log U$ coefficient (\ref{treeSGlogU}). Quite remarkably, in the examples we have checked we find the CFT expectation (\ref{treeSGlogU}) is always met with just the ansatz (\ref{ansatz22pp}), and therefore {\it no} single poles of the form (\ref{singlepole}) are needed.

\subsubsection*{The algorithm}
Let us now summarize the above strategy into the following concrete algorithm for computing the one-loop Mellin amplitude $\widetilde{\mathcal{M}}_{22pp}$
\begin{enumerate}
\item We compute the leading logarithmic singularities $\mathcal{H}^{(2)}_{22pp}\big|_{\log^2U}$, $\mathcal{H}^{(2)}_{2pp2}\big|_{\log^2U}$ and $\mathcal{H}^{(2)}_{2p2p}\big|_{\log^2U}$ from (\ref{lls}).
\item We extract the $\log^2V$ coefficients from the leading logarithmic singularities, and compare them with the $\log^2U\log^2V$ coefficients from the contour integrations for $\widetilde{\mathcal{M}}^{(2)}_{22pp}$, $\widetilde{\mathcal{M}}^{(2)}_{2pp2}$, $\widetilde{\mathcal{M}}^{(2)}_{2p2p}$. This fixes the parameters $c^s_{mn}$, $c^u_{mn}$, $c^t_{mn}$ for values of $m$ and $n$ that give rise to simultaneous triple poles in the ansatz.
\item In the small $V$ expansion, we extract the leading $\log V$ coefficients from the logarithmic singularities. They allow us to fix the rest of the simultaneous poles in the ansatz which give at most $\log^2U\log V$ singularities.
\item We resum the Mellin amplitudes, and compute the full $V$-dependence of the $\log^2U$ coefficients. We check that the fixed ansatz completely reproduces the leading logarithmic singularities as a function of $V$.  
\item We take the residues at $s=4,\ldots 2p-2$ in the inverse Mellin transformation with $\widetilde{\mathcal{M}}^{(2)}_{22pp}$. We focus on the leading $\log U$ coefficients and compute their full $V$-dependence. We check that they agree with the CFT expectation of (\ref{treeSGlogU}), and exclude the possibility of adding single poles.
\end{enumerate}

\subsection{Explicit examples}\label{subsecexamples}
We now apply the above algorithm to compute a few low-lying cases. We find that in all these examples the simultaneous poles are sufficient, and no single poles are needed.

\subsubsection*{\underline{$\langle \mathcal{O}_2^{SG}\mathcal{O}_2^{SG}\mathcal{O}_2^{SG}\mathcal{O}_2^{SG}\rangle$}}
The $p=2$ case was computed in \cite{Alday:2018kkw}. Here we just record the solution. Because of the enhanced crossing symmetry, the simultaneous pole coefficients are equal in all three channel. We have 
\begin{equation}
c^s_{mn}=c^t_{mn}=c^u_{mn}=\frac{P^{(6)}_{2,mn}}{(m+n-1)_5}
\end{equation}
where $p^{(6)}_{2,mn}$ is a degree 6 symmetric polynomial
\begin{equation}
\begin{split}
P^{(6)}_{2,mn}={}&\frac{32}{5}\big(-40 m - 8 m^2 + 36 m^3 + 12 m^4 - 40 n - 76 m n + 77 m^2 n + 
 114 m^3 n + 25 m^4 n\\
 {}& - 8 n^2 + 77 m n^2 + 216 m^2 n^2 + 
 120 m^3 n^2 + 15 m^4 n^2 + 36 n^3 + 114 m n^3 + 120 m^2 n^3 \\
 {}&+ 
 30 m^3 n^3 + 12 n^4 + 25 m n^4 + 15 m^2 n^4\big)\;.
\end{split}
\end{equation}
It was checked that the leading logarithmic singularity $\mathcal{H}^{(2)}_{2222}$ is fully reproduced as a function of $V$ by this ansatz. We also note that in this example there is no region in which single poles can be added.  

\subsubsection*{\underline{$\langle \mathcal{O}_2^{SG}\mathcal{O}_2^{SG}\mathcal{O}_3^{SG}\mathcal{O}_3^{SG}\rangle$}}
The $p=3$ case was computed in position space in \cite{Aprile:2017qoy}. Here we give a Mellin space derivation of their result. We start from $\mathcal{H}^{(2)}_{2233}$ in the s-channel. From the $\log^2U\log^2V$ coefficients we find  
\begin{equation}\label{cumnpeq3}
c^u_{mn}=\frac{P^{(6)}_{3,mn}}{(m+n-1)_5}
\end{equation}
with 
\begin{equation}
\begin{split}
P^{(6)}_{3,mn}={}&16\big(21 m^4 n^2+39 m^4 n+20 m^4+42 m^3 n^3+174 m^3 n^2+184 m^3 n+64 m^3+21 m^2 n^4\\
{}&+162 m^2 n^3+302 m^2 n^2+127 m^2 n-8 m^2+27 m n^4+120 m n^3+63 m n^2-130 m n\\
{}&-76 m+12 n^4+36 n^3-8 n^2-40 n\big)
\end{split}
\end{equation}
when $m\geq1$. The $m=0$ case does not contribute to the $\log^2U\log^2V$ coefficient, and therefore cannot be determined in this way.

 We now look at $\mathcal{H}^{(2)}_{2332}$ in the s-channel, and the relevant poles are 
 \begin{equation}
 \sum_{m,n=0}^\infty \frac{c^t_{mn}}{(s-5-2n)(t-4-2m)}\;.
 \end{equation} 
 All $m\geq 1$, $n\geq 0$ give rise to triple poles in the inverse Mellin integrand, and therefore the residues $c^t_{mn}$ can be fixed from the $\log^2U\log^2V$ coefficient of $\mathcal{H}^{(s)}_{2332}$. We find the coefficients $c^t_{mn}$ are given by the same formula
\begin{equation}\label{ctmnpeq3}
c^t_{mn}=\frac{P^{(6)}_{3,mn}}{(m+n-1)_5}\;,
\end{equation}
as we have expected from crossing symmetry. Moreover, by focusing on the $V^{-1}\log V$ coefficient in the small $V$ expansion of $\mathcal{H}^{(s)}_{2332}\big|_{\log^2U}$, we can fix the remaining $c^t_{mn}$ coefficients with $m=0$. These coefficients also fit into the expression (\ref{ctmnpeq3}). However a word of caution is needed. The formula (\ref{ctmnpeq3}) is {\it ambiguous} at $(m,n)=(0,0),(0,1)$, and gives different answers depending on the order we evaluate $m$ and $n$ in (\ref{ctmnpeq3}). The ambiguity follows from the Pochhammer denominator which vanishes at these values. Note that in matching the $\log^2U$ coefficient of $\mathcal{H}^{(s)}_{2332}$, we keep $m$ fixed at different values (corresponding to different powers of $U$) when computing for arbitrary $n$ (including $n=0$ that corresponds to $V^{-1}\log V$). This defines the correct order in evaluating (\ref{ctmnpeq3}), namely, we first evaluate $m$ and then evaluate $n$. By crossing symmetry,  the formula also extends to $m=0$ for $c^u_{mn}$ once the correct ordering is taken.

Finally, repeating the analysis for $\mathcal{H}^{(2)}_{2323}$ in the s-channel yields the coefficients $c^s_{mn}$ 
\begin{equation}\label{csmnpeq3}
c^s_{mn}=\frac{Q^{(6)}_{3,mn}}{(m+n-1)_5}\;,
\end{equation}
where $Q^{(6)}_{3,mn}$ is a symmetric polynomial 
\begin{equation}
\begin{split}
Q^{(6)}_{3,mn}={}&16\big(21 m^4 n^2+39 m^4 n+20 m^4+42 m^3 n^3+168 m^3 n^2+170 m^3 n+56 m^3+21 m^2 n^4\\
{}&+168 m^2 n^3+300 m^2 n^2+111 m^2 n-16 m^2+39 m n^4+170 m n^3+111 m n^2-104 m n\\
{}&-60 m+20 n^4+56 n^3-16 n^2-60 n\big)\;.
\end{split}
\end{equation}
This formula applies for all $m\geq0$, $n\geq 0$. Moreover, one can check that (\ref{csmnpeq3}) does not have the ordering ambiguity in this range.

Inserting the above coefficients into the ansatz, it is straightforward to check that the full logarithmic singularities are reproduced in three channels. To further eliminate the possibility of adding single poles at $s=4$, we can integrate the ansatz and obtain the leading $\log U$ coefficient. We find the ansatz gives
\begin{equation}
\mathcal{H}^{(2)}_{2233}\big|_{\log U}= U^4\left(\frac{1440 (V+1) \log (V)}{(V-1)^5}-\frac{480 \left(V^2+4 V+1\right) \log ^2(V)}{(V-1)^6}\right)+\mathcal{O}(U^5)
\end{equation}
which translates into the conformal block decomposition 
\begin{equation}
\sum_{\ell=0}^\infty-\frac{15 \sqrt{\pi } 2^{-2 \ell } \left((-1)^{\ell }+1\right) \Gamma (\ell +4)}{(\ell +1) (\ell +6) \Gamma \left(\ell +\frac{7}{2}\right)}g^{0,0}_{8+\ell,\ell}(z,\bar{z})+\mathcal{O}(U^5)\;.
\end{equation}
This should be compared with the twist 4 long operator coefficients at tree level
\begin{equation}
\mathcal{H}^{(1),long}_{2233}=\sum_{\ell=0}^\infty \frac{5 \sqrt{\pi } 2^{-2 \ell -4} \left((-1)^{\ell }+1\right) \Gamma (\ell +4)}{\Gamma \left(\ell +\frac{7}{2}\right)}g^{0,0}_{8+\ell,\ell}(z,\bar{z})+\mathcal{O}(U^5)\;.
\end{equation}
The decomposition coefficients are different by precisely 
\begin{equation}
-\frac{48}{(\ell +1) (\ell +6)}=\frac{1}{2}\gamma^{(1)}_{\tau=4,\ell,1}\;.
\end{equation}
Since this already agrees with the CFT expectation, no additional single poles can be added.

\subsubsection*{\underline{$\langle \mathcal{O}_2^{SG}\mathcal{O}_2^{SG}\mathcal{O}_4^{SG}\mathcal{O}_4^{SG}\rangle$}}
The analysis for $p=4$ is similar to $p=3$. From $\mathcal{H}^{(2)}_{2244}\big|_{\log^2U \log^2V}$, we obtain the coefficients 
\begin{equation}\label{cumnpeq4}
c^u_{mn}=\frac{P^{(6)}_{4,mn}}{(m+n-1)_5}
\end{equation}
for $m\geq 2$, with 
\begin{equation}
\begin{split}
P^{(6)}_{3,mn}={}&32\big(14 m^4 n^2+28 m^4 n+15 m^4+28 m^3 n^3+119 m^3 n^2+135 m^3 n+50 m^3+14 m^2 n^4\\
{}&+105 m^2 n^3+200 m^2 n^2+93 m^2 n-3 m^2+14 m n^4+59 m n^3+13 m n^2-106 m n\\
{}&-62 m+6 n^4+18 n^3-4 n^2-20 n\big)\;.
\end{split}
\end{equation}

From $\mathcal{H}^{(2)}_{2442}\big|_{\log^2U \log^2V}$, we recover the same formula for $m\geq 2$, $n\geq 0$, and find
\begin{equation}
c^t_{mn}=c^u_{mn}\;.
\end{equation}
The $c^t_{mn}$ coefficients with $m=0,1$ are found by comparing with the $V^{-2}\log V$ and $V^{-1}\log V$ coefficient of $\mathcal{H}^{(2)}_{2442}\big|_{\log^2U}$, and fit in the formula (\ref{cumnpeq4}).
Similar to the $p=3$ case, $c^t_{mn}$ is ambiguous for $(m,n)=(0,0),(0,1),(1,0)$. Comparison with $\mathcal{H}^{(2)}_{2442}\big|_{\log^2U}$ shows the correct prescription is again to first evaluate $m$ then evaluate $n$. 

From $\mathcal{H}^{(2)}_{2424}\big|_{\log^2U \log^2V}$, we obtain the remaining coefficients 
\begin{equation}\label{csmnpeq4}
c^s_{mn}=\frac{Q^{(6)}_{4,mn}}{(m+n-1)_5}\;,
\end{equation}
where $Q^{(6)}_{4,mn}$ is symmetric 
\begin{equation}
\begin{split}
Q^{(6)}_{4,mn}={}&32\big(14 m^4 n^2+28 m^4 n+15 m^4+28 m^3 n^3+112 m^3 n^2+118 m^3 n+40 m^3+14 m^2 n^4\\
{}&+112 m^2 n^3+200 m^2 n^2+76 m^2 n-13 m^2+28 m n^4+118 m n^3+76 m n^2-68 m n\\
{}&-42 m+15 n^4+40 n^3-13 n^2-42 n\big)\;,
\end{split}
\end{equation}
and free of ambiguities for $m\geq0$, $n\geq 0$.

These coefficients guarantee that all the leading logarithmic singularities $\mathcal{H}^{(2)}_{2244}\big|_{\log^2U}$, $\mathcal{H}^{(2)}_{2442}\big|_{\log^2U}$, $\mathcal{H}^{(2)}_{2424}\big|_{\log^2U}$ are reproduced. We also evaluate the residues at $s=4$ and $s=6$, which gives leading $\log U$ coefficients in the small $U$ expansion. They yield the following equations for the CFT data
\begin{equation}\label{2244logUeqns}\small
C^{(0)}_{22K_{4,\ell,1}}\gamma^{(1)}_{4,\ell,1}C^{(1)}_{44K_{4,\ell,1}}=-\frac{30 \sqrt{\pi } 2^{1-2 \ell } \Gamma (\ell +4)}{(\ell +1) (\ell +6) \Gamma \left(\ell +\frac{7}{2}\right)}\;,
\end{equation} 
\begin{equation}\small
C^{(0)}_{22K_{6,\ell,1}}\gamma^{(1)}_{6,\ell,1}C^{(1)}_{44K_{6,\ell,1}}+C^{(0)}_{22K_{6,\ell,2}}\gamma^{(1)}_{6,\ell,1}C^{(1)}_{44K_{6,\ell,2}}=\frac{90 \sqrt{\pi } 2^{-2 (\ell +1)} (\ell +3) (\ell +4) (19 \ell  (\ell +9)+446) \Gamma (\ell +1)}{(\ell +7) (\ell +8) \Gamma \left(\ell +\frac{9}{2}\right)}\;.
\end{equation} 
From the tree-level correlator, we have the conditions 
\begin{equation}\small
C^{(0)}_{22K_{4,\ell,1}}C^{(1)}_{44K_{4,\ell,1}}=\frac{5 \sqrt{\pi } 2^{-2 \ell -3} \Gamma (\ell +4)}{\Gamma \left(\ell +\frac{7}{2}\right)}\;,
\end{equation} 
\begin{equation}\small
C^{(0)}_{22K_{6,\ell,1}}C^{(1)}_{44K_{6,\ell,1}}+C^{(0)}_{22K_{6,\ell,2}}C^{(1)}_{44K_{6,\ell,2}}=-\frac{21 \sqrt{\pi } 2^{-2 (\ell +3)} \Gamma (\ell +5)}{\Gamma \left(\ell +\frac{9}{2}\right)}\;,
\end{equation} 
\begin{equation}\small
C^{(0)}_{33K_{6,\ell,1}}C^{(1)}_{44K_{6,\ell,1}}+C^{(0)}_{33K_{6,\ell,2}}C^{(1)}_{44K_{6,\ell,2}}=\frac{3 \sqrt{\pi } 2^{-2 (\ell +3)} (\ell  (\ell +9)+29) \Gamma (\ell +5)}{\Gamma \left(\ell +\frac{9}{2}\right)}\;.
\end{equation} 
Using the results \cite{Aprile:2017xsp,Aprile:2017qoy}\footnote{Note that OPE coefficients are different by some overall factors because we normalize the external operators to have unit two-point functions.}
\begin{equation}\small
\begin{split}
{}&C^{(0)}_{22K_{4,\ell,1}}=\sqrt{\frac{(\ell +1) (\ell +6) \Gamma (\ell +4)^2}{3\Gamma (2 \ell +7)}}\;,\quad C^{(0)}_{22K_{6,\ell,1}}=\sqrt{\frac{(\ell +1) (\ell +2) (\ell +8) \Gamma (\ell +5)^2}{10\Gamma (2 (\ell +5))}}\;,\\
{}& C^{(0)}_{22K_{6,\ell,2}}=\sqrt{\frac{(\ell +1) (\ell +7) (\ell +8) \Gamma (\ell +5)^2}{10\Gamma (2 (\ell +5))}}\;, \quad  C^{(0)}_{33K_{6,\ell,1}}=-\frac{(\ell+7)}{2}\sqrt{\frac{(\ell +1) (\ell +2) (\ell +8) \Gamma (\ell +5)^2}{10\Gamma (2 (\ell +5))}}\\
{}&C^{(0)}_{33K_{6,\ell,2}}=\frac{(\ell+2)}{2}\sqrt{\frac{(\ell +1) (\ell +7) (\ell +8) \Gamma (\ell +5)^2}{10\Gamma (2 (\ell +5))}}\;,
\end{split}
\end{equation}
we find 
\begin{equation}\small
\begin{split}
{}&C^{(1)}_{44K_{4,\ell,1}}=40\sqrt{\frac{3 \Gamma (\ell +4)^2}{(\ell +1) (\ell +6) \Gamma (2 \ell +7)}}\;,\\
{}&C^{(1)}_{44K_{6,\ell,1}}=-12\sqrt{\frac{10 (\ell +8) (2 \ell +9) \Gamma (\ell +5)^2}{(\ell +1) (\ell +2) \Gamma (2 \ell +9)}}\;, \\
{}&C^{(1)}_{44K_{6,\ell,2}}=12\sqrt{\frac{10 (\ell +1) (2 \ell +9) \Gamma (\ell +5)^2}{(\ell +7) (\ell +8) \Gamma (2 \ell +9)}}\;.
\end{split}
\end{equation}
Inserting the solution into (\ref{2244logUeqns}), and using \cite{Aprile:2017xsp}
\begin{equation}
\gamma^{(1)}_{6,\ell,1}=-\frac{480}{(\ell +1) (\ell +2)}\;,\quad \gamma^{(1)}_{6,\ell,2}=-\frac{480}{(\ell +7) (\ell +8)}\;,
\end{equation}
we find the equations are satisfied. This concludes that the Mellin amplitudes  for $p=4$ does not allow single poles. This pattern of no single poles should be a general feature, and we conjecture that it is true for any $p$.

\subsection{Pre-amplitudes and general Mellin amplitudes of $\langle \mathcal{O}_2^{SG}\mathcal{O}_2^{SG}\mathcal{O}_p^{SG}\mathcal{O}_p^{SG}\rangle$}\label{subsecpreampli}
As was demonstrated in the previous section, the algorithm outlined in Section \ref{subsec22pp} gives a concrete way to compute case-by-case one-loop Mellin amplitudes for  $\langle \mathcal{O}_2^{SG}\mathcal{O}_2^{SG}\mathcal{O}_p^{SG}\mathcal{O}_p^{SG}\rangle$. While the residue coefficients $c^{s,t,u}_{mn}$ clearly share the same structures for different values of $p$, their precise $p$-dependence is not immediately obvious from these examples. It would be nice to write down these coefficients in a closed form for arbitrary $p$, and understand the origin of the structural similarity in all the cases. In this section, we achieve this by reducing the Mellin amplitudes to the {\it pre-amplitudes}, which manifests the implication of the hidden conformal symmetry.

Let us recall the hidden symmetry relation (\ref{lls}). The leading logarithmic singularity of the one-loop reduced correlator is computed by the action of $\Delta^{(8)}$ on an auxiliary object which we denote as $\lambda_{p_1p_2p_3p_4}\big|_{\log^2U}$ 
\begin{equation}
\mathcal{H}^{(2)}_{p_1p_2p_3p_4}\big|_{\log^2U}=\Delta^{(8)}\lambda_{p_1p_2p_3p_4}\big|_{\log^2U}\;,
\end{equation}
and this object is further related to a ``seed'' via  
\begin{equation}\label{diffrelambdalog2U}
\lambda_{p_1p_2p_3p_4}\big|_{\log^2U}=\mathcal{D}_{p_1p_2p_3p_4}\lambda_{2222}\big|_{\log^2U}\;.
\end{equation}
From the notation it is clear that it is natural to propose the existence of a function $\lambda_{p_1p_2p_3p_4}$, which enjoys the same hidden conformal symmetry as the tree level reduced correlator $\mathcal{H}^{(1)}_{p_1p_2p_3p_4}$
\begin{equation}\label{diffrelambda}
\lambda_{p_1p_2p_3p_4}=\mathcal{D}_{p_1p_2p_3p_4}\lambda_{2222}\;.
\end{equation}
The relation (\ref{diffrelambdalog2U}) then follows as a consequence. 
Note that the differential operator $\Delta^{(8)}$ is purely kinematical. We can therefore expect simplification if we factor out $\Delta^{(8)}$ and study the Mellin transformation of $\lambda_{p_1p_2p_3p_4}$, which we will call the {\it pre-amplitude}. The Mellin representation was defined in  (\ref{lambdaMellin}), and the pre-amplitude was denoted as $\mathcal{L}_{p_1p_2p_3p_4}$.

On the other hand, since only information about $\lambda_{p_1p_2p_3p_4}\big|_{\log^2U}$ is available, we cannot completely fix $\mathcal{L}_{p_1p_2p_3p_4}$. In fact, the limited information only allows us to obtain the poles in $s$ which overlap with the Gamma functions to form triple poles. Nevertheless, as we will see, we can already get a lot of mileage from this partial information.

Let us start with the simplest case where $k_i=2$. We use $\lambda_{2222}\big|_{\log^2U}$ as input. Applying the same techniques for computing $\widetilde{\mathcal{M}}^{(2)}_{p_1p_2p_3p_4}$ we can fix partially the pre-amplitude $\mathcal{L}_{p_1p_2p_3p_4}$. We find that $\mathcal{L}_{2222}$ must have the following singularities in order to  reproduce the full $V$-dependence in $\lambda_{2222}\big|_{\log^2U}$
\begin{equation}
\mathcal{L}_{2222}=\sum_{n,m=0}^\infty \left(\frac{b_{mn}}{(s-4-2m)(t-4-2n)}+\frac{b_{mn}}{(s-4-2m)(u-4-2n)}\right)+\mathcal{P}_{2222}\;.
\end{equation}
Here $s+t+u=6$, and the coefficients $b_{mn}$ are determined to be
\begin{equation}
b_{mn}=\frac{4 (n-1) n}{15 (m+1) (m+2) (m+n+1) (m+n+2) (m+n+3)}
\end{equation} 
The function $\mathcal{P}_{2222}$ contains additional terms in $\mathcal{L}_{2222}$ which do not contribute to the coefficient of $\log^2U$, and is in general singular.

Similar analysis can be done for $\mathcal{L}_{p_1p_2p_3p_4}$. However, a more convenient way to obtain the pre-amplitudes is to exploit the relation (\ref{diffrelambda}) in Mellin space.  It yields the following difference relation
\begin{equation}\label{Ldifferencere}
\mathcal{L}_{p_1p_2p_3p_4}=\widehat{\mathcal{D}}_{p_1p_2p_3p_4}\circ \mathcal{L}_{2222}\;.
\end{equation}
where the difference operator $\widehat{\mathcal{D}}_{p_1p_2p_3p_4}$ already appeared at tree-level for the Mellin amplitudes $\widetilde{\mathcal{M}}^{(1)}_{p_1p_2p_3p_4}$
\begin{equation}
\widetilde{\mathcal{M}}^{(1)}_{p_1p_2p_3p_4}=\widehat{\mathcal{D}}_{p_1p_2p_3p_4}\circ\widetilde{\mathcal{M}}^{(1)}_{2222}\;.
\end{equation}
 The case of $\widehat{\mathcal{D}}_{22pp}$ is particularly simple, and we find that $\mathcal{L}_{22pp}$ must contain singularities 
\begin{equation}\label{L22pp}
\begin{split}
\mathcal{L}_{22pp}={}&\frac{p}{2 \Gamma (p-1)}\bigg[\sum_{n,m=0}^\infty \frac{b_{mn}}{(s-4-2m)(t-(p+2)-2n)}\\
{}&\quad\quad\quad\quad\quad\quad\quad\quad\quad+\sum_{n,m=0}^\infty\frac{b_{mn}}{(s-4-2m)(u-(p+2)-2n)}\bigg]+\mathcal{P}_{22pp}
\end{split}
\end{equation}
where $s+t+u=2p$. Note that when $m\geq p-2$ the inverse Mellin integrand has triple poles, and we can cross check that this formula agrees with the position space calculation by applying $\mathcal{D}_{22pp}$ on $\lambda_{2222}\big|_{\log^2U}$.  

We still need to relate the pre-amplitudes to the Mellin amplitudes. This is achieved by translating the differential operator $\Delta^{(8)}$ into a difference operator $\widehat{\Delta}^{(8)}$ in Mellin space. The explicit form of $\widehat{\Delta}^{(8)}$ is quite cumbersome, which we will refrain from writing down. However it is rather straightforward to derive this operator, and we sketch the derivation below. We start from the inverse Mellin integral for $\lambda_{22pp}$
\begin{equation}
\lambda_{22pp}=\int \frac{dsdt}{(4\pi i)^2} U^{\frac{s}{2}+2}V^{\frac{t-2-p}{2}}\mathcal{L}_{22pp}(s,t)\Gamma[\tfrac{4-s}{2}]\Gamma[\tfrac{2p-s}{2}]\Gamma^2[\tfrac{2+p-t}{2}]\Gamma^2[\tfrac{2+p-u}{2}]\;,
\end{equation}
and act with $\Delta^{(8)}$. We can use a change of variables to write $\Delta^{(8)}$ in terms of $U$ and $V$,\footnote{In this case the R-symmetry dependence is trivial because $\mathcal{L}_{22pp}$ contains only one irreducible representation $[0,0,0]$. See discussion in Appendix \ref{app:diffops} for details.}   which acts on the $U^{\frac{s}{2}+2}V^{\frac{t-2-p}{2}}$ factor in elementary ways.\footnote{Equivalently, we can use the change of variables (\ref{defUV}) to write $U^{\frac{s}{2}+2}V^{\frac{t-2-p}{2}}$ in terms of $z$, $\bar{z}$. The action of $\Delta^{(8)}$ on this expression is then very simple, and the result can be easily rewritten in terms of $U$ and $V$.} The resulting expression includes powers of $U$ and $V$ in addition to the factor $U^{\frac{s}{2}+2}V^{\frac{t-2-p}{2}}$, but these powers can be absorbed by shifting $s$ and $t$. Finally, we bring together all the terms into the same form as the Mellin representation for $\mathcal{H}^{(2)}_{22pp}$
\begin{equation}
\mathcal{H}^{(2)}_{22pp}=\int \frac{dsdt}{(4\pi i)^2} U^{\frac{s}{2}+2}V^{\frac{t-2-p}{2}}\widetilde{\mathcal{M}}^{(2)}_{22pp}(s,t)\Gamma[\tfrac{4-s}{2}]\Gamma[\tfrac{2p-s}{2}]\Gamma^2[\tfrac{2+p-t}{2}]\Gamma^2[\tfrac{2+p-u}{2}]\;,
\end{equation}
This defines a difference operator $\widehat{\Delta}^{(8)}$ which relates $\widetilde{\mathcal{M}}^{(2)}_{22pp}$ and $\mathcal{L}_{22pp}$
\begin{equation}\label{M22ppviaL22pp}
\widetilde{\mathcal{M}}^{(2)}_{22pp}=\widehat{\Delta}^{(8)}\circ \mathcal{L}_{22pp}\;.
\end{equation}

 We can now obtain the simultaneous pole coefficients $c^{t,u}_{mn}$ by using the relation (\ref{M22ppviaL22pp}). Focusing on a particular pair of simultaneous poles labelled by $m$ and $n$, we can read off its coefficient 
\begin{equation}\label{ctumngenp}
c^{t,u}_{mn}=\frac{P^{(6)}_{p,mn}}{(m+n-1)_5}
\end{equation}
where the numerator polynomial is given by 
\begin{equation}\small
\begin{split}
{}&P^{(6)}_{p,mn}=\frac{2p (p+1) (p+2)}{15 \Gamma (p-1)}\bigg\{(m-1) m (n+1) (n+2)(m+n+2) (m+n+3) p^2 \\
{}& \quad\quad+m (1 + n) (3 + m + n) m^2 (6 + 7 n)p+m (1 + n) (3 + m + n) (2 + n) (-2 + 9 n)p\\
{}&\quad\quad+m (1 + n) (3 + m + n) (-2 + 17 n + 7 n^2)p+(-1 + m) m (1 + n) (3 + m + n) (-2 + 17 n + 7 n^2) p\\
{}&\quad\quad+4 (1 + m) (2 + m) (-1 + n) n (10 + 12 n + 3 n^2)+4 m (1 + m) (2 + m) (-1 + 2 n + 12 n^2 + 6 n^3)\\
{}&\quad\quad+4 (1 + m) (2 + m) m^2 (1 + 3 n + 3 n^2)\bigg\}\;.
\end{split}
\end{equation}
This formula reproduces the cases with $p=2,3,4$ which we have computed explicitly. Moreover, the formula also has ambiguities for $(m,n)=(0,0),(0,1),(1,0)$, and should be accompanied by the prescription that we evaluate $m$ before $n$. 

We would like to point out that  (\ref{ctumngenp}) was derived by assuming $\{m,n\}$ are in the ``bulk'', {\it i.e.}, when all the simultaneous poles labelled by $\{m',n'\}$ in $\mathcal{L}_{22pp}$,  requested by $\widehat{\Delta}^{(8)}$ to reproduce certain simultaneous poles in $\widetilde{\mathcal{M}}^{(2)}_{22pp}$ labelled by $\{m,n\}$, are present. The relation  (\ref{ctumngenp}) is known to fail near the boundary of $\{m,n\}$. We suspect that it has to do with the subtleties of the contours, which were ignored in deriving the differential operator. Moreover, $\mathcal{L}_{22pp}$ may contain other singularities which do not contribute to $\log^2U$ but play a role near the boundary. We plan to perform a more detailed study about such effects in the future, and will not further dwell on this issue here. Nevertheless, as a prescription to obtain a general formula for $c^{t,u}_{mn}$, our procedure appears always to be valid. 

An analogous analysis can be performed for $\mathcal{H}^{(2)}_{2p2p}$, where $\mathcal{L}_{2p2p}$ contains the simultaneous poles 
\begin{equation}\label{L2p2p}\small
\frac{p}{2 \Gamma (p-1)}\bigg[\sum_{n,m=0}^\infty \frac{b_{mn}}{(s-(p+2)-2m)(t-(p+2)-2n)}+\frac{b_{mn}}{(s-(p+2)-2m)(u-4-2n)}\bigg]\;.
\end{equation}
We can similarly derive the difference operator $\widehat{\Delta}^{(8)}$ which takes a slightly different form. Using the difference  relation between $\mathcal{L}_{2p2p}$ and $\widetilde{\mathcal{M}}^{(2)}_{2p2p}$, we obtain the following expression for $c^s_{mn}$
\begin{equation}
c^s_{mn}=\frac{Q^{(6)}_{p,mn}}{(m+n-1)_5}\;,
\end{equation}
with 
\begin{equation}\small
\begin{split}
{}&Q^{(6)}_{p,mn}=\frac{2p (p+1) (p+2)}{15 \Gamma (p-1)}\bigg\{(1 + m) (2 + m) (1 + n) (2 + n) (-1 + m + n) (m + n)p^2\\
{}&\quad\quad\quad+2 (m+1) (n+1) n (3 n+10) (m+n-1)p+(m+1) m^2 (n+1) (7 n+6) (m+n-1)p\\
{}&\quad\quad\quad+(m+1) m (n+1) \left(7 n^2+36 n+20\right) (m+n-1)p+4 (n-1) n (m+n+2) (m+n+3)\\
{}&\quad\quad\quad+12 m^2 n (n+1) (m+n+2) (m+n+3)+4 m (m-1) (m+n+2) (m+n+3)\\
{}&\quad\quad\quad+4 m n (3 n-1) (m+n+2) (m+n+3)\bigg\}\;.
\end{split}
\end{equation}
When plugging in $p=2,3,4$, we reproduce the previous results.

\subsection{Comments on higher-weight correlators}\label{subsechigherweight}
Compared to $\langle O^{SG}_2 O^{SG}_2 O^{SG}_p O^{SG}_p\rangle$, a few new features emerge in a general correlator:
\begin{itemize}
\item In the s-channel of $\langle O^{SG}_2 O^{SG}_2 O^{SG}_p O^{SG}_p\rangle$, we have seen that the double-trace spectra have a non-overlapping region in the s-channel, which has finite with support on the conformal twist. In the generic case, non-overlapping regions show up in all three channels.
\item The free correlators in the supergravity basis are no longer $1/c$-exact (see, {\it e.g.}, \cite{Alday:2014qfa} for the explicit example of $\langle O^{SG}_3 O^{SG}_3 O^{SG}_3 O^{SG}_3\rangle_{free}$).
\item The reduced correlator can have nontrivial R-symmetry dependence. In general, it is a polynomial in the R-symmetry cross ratios of degree $\min\{p_4,\frac{p_2+p_3+p_4-p_1}{2}\}-2$ (assuming $p_1\geq p_2\geq p_3\geq p_4$).
\end{itemize}
Let us comment on the consequence of some of these properties on the analytic structure of the Mellin amplitude.

To begin, we notice that when the free correlator does not truncate at $\mathcal{O}(1/c)$, spurious long operator may contribute to $\widetilde{H}^{(2)}_{p_1p_2p_3p_4}$. When this happens, the one-loop Mellin amplitude consequently must have poles to cancel these spurious contributions, and such poles will reside in the ``tree-level region''
\begin{equation}
B_{tree}=\left\{(Re(s),Re(t))\bigg|\begin{tabular}{c}$Re(s)<\min\{p_1+p_2,p_3+p_4\}$ \\$Re(t)<\min\{p_1+p_4,p_2+p_3\}$ \\$Re(u)<\min\{p_1+p_3,p_2+p_4\}$\end{tabular}\right\}
\end{equation}
where all the poles in tree level supergravity solution (\ref{treeMellin}) sit.

However, unlike in the t-channel of $\langle O^{SG}_2 O^{SG}_2 O^{SG}_p O^{SG}_p\rangle$ where only poles above the double-trace minimum correspond to long physical operators, poles inside the tree-level region $B_{tree}$ can become physical too. A simple example that illustrates this point is the singlet twist 4 long operators in the one-loop correlator $\langle O^{SG}_3 O^{SG}_3 O^{SG}_3 O^{SG}_3\rangle$. To see such operators are indeed exchanged at one loop, we note that twist 4 long operators $[O_2O_2]_{0,\ell}$ first appear in $\langle O^{SG}_2 O^{SG}_2 O^{SG}_3 O^{SG}_3\rangle$ at tree level with $\mathcal{O}(1/c)$ coefficients. The squared OPE coefficients therefore have the correct power in $1/c$.

These new features make solving the Mellin amplitudes for operators with general weights a more challenging task. Nevertheless, it is conceivable that the structure of simultaneous poles should persist in the generic case, at least for the poles which overlap with both Gamma functions in a given channel. The coefficients of the simultaneous poles are now polynomials in the R-symmetry cross ratios, but we expect that they do not depend on the Mandelstam variables. Our strategy from Section \ref{subsec22pp} can then be applied straightforwardly to fix the coefficients of these simultaneous poles, and determines a large part of the Mellin amplitude. However, fixing the low-lying poles must await a further systematic study of the tree-level correlators. In particular, OPE coefficients of the type $C^{(1)}_{pp K_{\tau,\ell,i}}$ with $\tau<2p$ should first be extracted, which are so far elusive in the literature. We will leave these problems for the future.

\section{Flat Space Limit}\label{secflatspace}
In this section we discuss the flat space limit of the Mellin amplitude found in the previous section. The details of the limit in this setting were discussed in \cite{Alday:2018pdi}, following \cite{Chester:2018dga}, to where we refer the reader for the details. Because we are considering the four-point correlator of KK modes with momentum on the internal $S^5$, one must adapt the formula by Penedones \cite{Penedones:2010ue} to this case. The first step is to reinstate the R-charge dependence to the ``reduced'' Mellin amplitude $\widetilde{\mathcal{M}}_{22pp}(s,t)$ and define the ``full'' Mellin amplitude $\mathcal{M}_{22pp}(s,t,\sigma,\tau)$. This is achieved by acting with the following difference operator $\widehat{R}$ \cite{Rastelli:2016nze,Rastelli:2017udc}:
\begin{equation}
\mathcal{M}_{22pp}(s,t,\sigma,\tau)= \widehat R \circ \widetilde{\mathcal{M}}_{22pp}(s,t)\;,
\end{equation}
with
\begin{equation}
\widehat R \equiv \tau + (1-\sigma-\tau)\widehat V + (\tau^2-\tau-\sigma \tau)\widehat U + (\sigma^2-\sigma-\sigma \tau)\widehat{UV}+\sigma \widehat{V^2} + \sigma \tau\widehat{U^2}
\end{equation}
where hatted powers act as 
\begin{align}
\widehat{U^mV^n}\circ \widetilde{\mathcal{M}}_{p_1p_2p_3p_4}(s,t) &\equiv \widetilde{\mathcal{M}}_{p_1p_2p_3p_4}(s-2m,t-2n; \sigma,\tau) \nonumber \\&\times \left(\frac{p_1+p_2-s}{2}\right)_m\left(\frac{p_3+p_4-s}{2}\right)_m\left(\frac{p_2+p_3-t}{2}\right)_n\\
&\times \left(\frac{p_1+p_4-t}{2}\right)_n\left(\frac{p_1+p_3-u}{2}\right)_{2-m-n}\left(\frac{p_2+p_4-u}{2}\right)_{2-m-n}\;, \nonumber
\end{align}
and in the case at hand $(p_1,p_2,p_3,p_4)=(2,2,p,p)$. In terms of the full Mellin amplitude, the flat space limit reads 
\begin{equation}
\label{flatlimit}
\lim_{L \to  \infty} L(L^5 V_5) \mathcal{M}_{22pp}(L^2 s, L^2 t,\sigma,\tau) = \frac{1}{\Gamma(p)} \int_0^\infty d\beta \beta^{p-1} e^{-\beta} {\cal A}_{p,\perp}^{(10d)}(2\beta s,2\beta t; \sigma,\tau)
\end{equation}
where $L$ is the common radius of $S^5$ and $AdS_5$ and $L^5 V_5 = \pi^3 L^5$ is the volume of the $S^5$. ${\cal A}_{p,\perp}^{(10d)}(2\beta s,2\beta t; \sigma,\tau)$ is the ten dimensional flat space amplitude of four supergravitons with momenta $k_i$ restricted to $AdS$, integrated against $S^5$ wavefunctions $\phi_2$ and $\phi_p$ and contracted with $SU(4)_R$ polarization vectors $t_i$, with transverse kinematics $k_i \cdot t_i =0$. 

Let us now consider the large $s,t$ limit of the Mellin amplitude. One can explicitly check that in this limit the action of the difference operator simply gives an overall factor
\begin{equation}
\lim_{s,t \to \infty}
\mathcal{M}_{22pp}(s, t,\sigma,\tau) = \frac{\Theta_4^{flat}(s,t;\sigma,\tau)}{16} \lim_{s,t \to \infty}\widetilde{\mathcal{M}}_{22pp}(s,t)\;,
\end{equation}
where
\begin{equation}
\Theta_4^{flat}(s,t;\sigma,\tau)= (t u + t s \sigma + s u \tau)^2\;,
\end{equation}
and remember that in the flat space limit $s+t+u=0$. Let us now consider the large $s,t$ limit of the reduced amplitude at one loop. In this case the amplitude takes the form
\begin{equation}
\widetilde{\mathcal{M}}^{(2)}_{22pp}(s,t)= \sum_{m,n=0}^\infty\frac{c_{mn}^u}{(s-4-2m)(t-(2+p)-2n)}+\cdots\;,
\end{equation}
where $c_{mn}^u$  was given in (\ref{ctumngenp}), and the dots represent the contribution from the other two channels. From now on we will ignore the other two channels, as they can be treated in exactly the same way. Taking the large $s,t$ limit of this sum is subtle, as one needs to perform the sums and then take the limit. Given the explicit form of  $c_{mn}^u$ this sum requires regularisation. A standard regularisation is zeta function regularisation, which in the present context is equivalent to take derivatives w.r.t. the Mellin variables $s,t$. Hence we consider instead
\begin{equation}
\partial_s \partial_t \widetilde{\mathcal{M}}^{(2)}_{22pp}(s,t)= \sum_{m,n=0}^\infty\frac{c_{mn}^u}{(s-4-2m)^2(t-(2+p)-2n)^2}\;.
\end{equation}
This is a hard sum, but convergent. A careful analysis of the sumand shows that the leading contribution in the large $s,t$ regime arises from the region $m,n \sim s,t$. In this limit furthermore we can replace the sum by an integral. From the explicit form of the coefficients $c_{mn}^u$ we obtain
\begin{equation}
\partial_s \partial_t \widetilde{\mathcal{M}}^{(2)}_{22pp}(s,t) = \frac{p(p+1)(p+2)(p+3)(p+4)}{\Gamma(p-1)} \partial_s \partial_t I(s,t)
\end{equation}
in the leading order in the large $s,t$ limit, where 
\begin{equation}
\partial_s \partial_t I(s,t) = \frac{2}{15} \int_0^\infty dm dn \frac{m^2 n^2}{(m+n)^3(s-2m)^2(t-2n)^2}\;.
\end{equation}
This is exactly the same integral encountered in \cite{Alday:2018kkw}. As explained there this integral can be performed and exactly agrees with the double derivative of the 10D box integral $I(s,t)$, with the correct prefactor. The comparison is done in the region where logarithms are real in the Euclidean region $s,t<0$. Explicitly we obtain
\begin{equation}
I(s,t) = \frac{1}{120} \Big( \frac{s^2t^2}{u^3} \left(\log^2\frac{-s}{-t}+\pi^2\right)-(s-t)\left(\frac{st}{u^2}+\frac12\right)\log\frac{-s}{-t}
+ u \log \frac{\sqrt{-s}\sqrt{-t}}{\Lambda^2}-\frac{st}{u}+c_1 \Lambda^2+c_2 u\Big)
\end{equation}
where $s+t+u=0$ and the constants $c_1,c_2$ encode the ambiguities present when using a regularisation scheme. We conclude 
\begin{equation}
\lim_{s,t \to \infty}\widetilde{\mathcal{M}}^{(2)}_{22pp}(s,t) = \frac{p(p+1)(p+2)(p+3)(p+4)}{\Gamma(p-1)} I(s,t)\;,
\end{equation}
so that in the flat space limit
\begin{equation}
\mathcal{M}^{(2)}_{22pp}(s, t,\sigma,\tau) = \frac{\Theta_4^{flat}(s,t;\sigma,\tau)}{16} \frac{p(p+1)(p+2)(p+3)(p+4)}{\Gamma(p-1)} I(s,t)\;.
\end{equation}
Does this reproduce the expected four graviton flat space amplitude? This amplitude admits the expansion
\begin{equation}
{\cal A}^{(10d)}(s,t; \sigma,\tau) = \hat K \left(-8 \pi G_N \frac{1}{s t u}+ \frac{(8\pi G_N)^2}{(4\pi)^5} \left( I(s,t) + \text{crossed} \right)+\cdots \right)
\end{equation}
where $G_n = \frac{\pi^4 L^8}{8 c}$ is the ten-dimensional Planck constant, and $\hat K$ is an overall kinematic factor. As already mentioned, in order to make a comparison we have to choose a specific polarisation for the gravitons and transverse kinematics $t_i \cdot k_i=0$. As shown in \cite{Alday:2018pdi} in this case $\hat K$ is proportional to $\Theta_4^{flat}(s,t;\sigma,\tau)$ and we obtain
\begin{equation}
{\cal A}_{p,\perp}^{(10d)}(s,t; \sigma,\tau) = c(p) \Theta_4^{flat}(s,t;\sigma,\tau) \left(-8 \pi G_N \frac{1}{s t u}+ \frac{(8\pi G_N)^2}{(4\pi)^5} \left( I(s,t) + \text{crossed} \right)+\cdots \right)
\end{equation}
where the coefficient $c(p)$ arises from the integration against the four KK-mode wave functions. Let us first consider the result at tree-level. Plugging ${\cal A}_{p,\perp}^{(10d)}(s,t; \sigma,\tau)$ at this order into (\ref{flatlimit}) and using 
\begin{equation}
\Theta_4^{flat}(2\beta s,2\beta t;\sigma,\tau) = 2^4 \beta^4 \Theta_4^{flat}(s,t;\sigma,\tau)\;,
\end{equation}
we obtain
\begin{equation}
 \frac{1}{\Gamma(p)} \int_0^\infty d\beta \beta^{p-1} e^{-\beta} {\cal A}_p^{(tree)}(2\beta s,2\beta t; \sigma,\tau) = -p c(p) \Theta_4^{flat}(s,t;\sigma,\tau) \frac{8 L^8 \pi^5}{s t u} a\;.
\end{equation}
According to (\ref{flatlimit}), this should reproduce the supergravity Mellin amplitude at tree level, in the large $s,t$ limit:
\begin{equation}
\widetilde{\mathcal{M}}^{(1)}_{22pp}(s,t)\sim - \frac{16p}{\Gamma(p-1)} \frac{1}{ s t u}
\end{equation}
where recall, this is the coefficient of $a=\frac{1}{4c}$. We see that we reproduce exactly the desired result provided
\begin{equation}
c(p)= \frac{1}{8\pi^2 \Gamma(p-1)}.
\end{equation}
Having fixed $c(p)$ we can now consider the problem at one-loop. Using $I(2\beta s, 2 \beta t) = 2\beta I(s,t)$, we obtain
\begin{eqnarray}
 &&\frac{1}{\Gamma(p)} \int_0^\infty d\beta \beta^{p-1} e^{-\beta} {\cal A}_p^{(one-loop)}(2\beta s,2\beta t; \sigma,\tau) = \\
 && \quad\quad\quad\quad\quad\quad\frac{\Theta_4^{flat}(s,t;\sigma,\tau)}{16} a^2 \left( \frac{p(p+1)(p+2)(p+3)(p+4)L^{16} \pi^3}{\Gamma(p-1)} I(s,t) + \text{crossed} \right)\;, \nonumber
\end{eqnarray}
which exactly agrees with the result for the Mellin amplitude in the flat space limit! The agreement with the flat space amplitude provides a nontrivial consistency check for our Mellin space formula. 

Before concluding this section let's make the following remark. As it is clear from this computation, the sum over poles that defines the Mellin amplitude is actually divergent, and needs to be regularised. This is a manifestation of the UV divergences also present in the flat space supergravity computation. Since UV divergences arise from small distance effects, we expect the structure of UV divergences in flat space and $AdS$ to be the same. Indeed, note that the regularisation procedure introduces an ambiguity that exactly corresponds to an $R^4$ term, which is the form of the counterterm present in the flat space computation.  

%%%%%%%%%%%
%%%%%%%%%%%
%%%%%%%%%%%
%%%%%%%%%%%

\section{Discussions and Outlook}\label{secconclusion}
In this paper we demonstrated the simplicity of $AdS_5\times S^5$ IIB supergravity  at one loop level by studying the $\langle \mathcal{O}_2^{SG}\mathcal{O}_2^{SG}\mathcal{O}_p^{SG}\mathcal{O}_p^{SG}\rangle$ correlators. We developed a systematic algorithm to construct the one-loop Mellin amplitudes (from the tree level data), and for $\langle \mathcal{O}_2^{SG}\mathcal{O}_2^{SG}\mathcal{O}_p^{SG}\mathcal{O}_p^{SG}\rangle$ we were able to obtain them in a closed form for any value of $p$. We also studied the large Mellin-Mandelstam variable limit of these amplitudes, and found a perfect agreement with the flat space expectation. 

The most evident feature of the one-loop Mellin amplitudes is their remarkably simple analytic structure. The amplitudes consist of only simultaneous poles in the Mellin-Mandelstam variables with constant coefficients, which are minimally required to reproduce the various logarithmic singularities that can appear at one loop. The fact that only simultaneous poles are involved is, in a way, reminiscent of the so-called {\it no-triangle property} of maximal supergravity. It is conceivable that the simplicity of the analytic structure will persist in the Mellin amplitudes of more general correlators, and we are optimistic that an elegant general solution can be found in Mellin space.

Another way to understand the analytic structure of the  $\langle \mathcal{O}_2^{SG}\mathcal{O}_2^{SG}\mathcal{O}_p^{SG}\mathcal{O}_p^{SG}\rangle$ correlators at one loop is that they are {\it completely} fixed by the hidden ten dimensional conformal symmetry, which determines the leading logarithmic singularities of the reduced correlators. It is important to emphasize, as we discussed in Section \ref{subsec22pp}, that there could {\it a priori} be single poles in the Mellin amplitudes which would destroy this feature. Remarkably, consistency conditions of CFT always rule out the existence of such single poles in the examples we studied. This seems to suggest that the implication of the hidden conformal symmetry extends beyond the leading logarithmic singularities. On the other hand, in general correlators there are more ways to modify the Mellin amplitude without changing the leading logarithmic singularities. It would be interesting to see to which extent the correlators are determined by the hidden conformal symmetry. More concretely, the pre-amplitudes and their associated correlators $\lambda_{p_1p_2p_3p_4}$ appear to be the right objects to focus on. In the case of $\langle \mathcal{O}_2^{SG}\mathcal{O}_2^{SG}\mathcal{O}_p^{SG}\mathcal{O}_p^{SG}\rangle$, we showed that the one-loop correlators are completely captured by the pre-amplitudes. The generalization to higher weights is straightforward, and it would be fascinating to see if the same still holds true. 

It is also clear what needs to be done in order to make progress, following our discussion in Section \ref{subsechigherweight}. The first order of business is to systematically extract all the OPE coefficients which make their first appearance at order $\mathcal{O}(1/c)$. These data will play an important role in fixing the one-loop Mellin amplitude. Acquiring these OPE coefficients demands a more thorough analysis of the tree-level correlators, and may reveal new structural features in the theory. The general correlators do not share the simplifications of the $\langle \mathcal{O}_2^{SG}\mathcal{O}_2^{SG}\mathcal{O}_p^{SG}\mathcal{O}_p^{SG}\rangle$ correlators. However, we can choose to study different classes of correlators so that only some of the new features are present. For example, we can simplify the analysis by first focusing on the next-to-next-to-extremal correlators\footnote{These are the correlators $\langle \mathcal{O}_{p_1}^{SG}\mathcal{O}_{p_1}^{SG}\mathcal{O}_{p_3}^{SG}\mathcal{O}_{p_4}^{SG}\rangle$ with $p_2+p_3+p_4-p_1=4$, assuming that $p_1$ is the largest weight.}, of which the reduced correlators are R-symmetry singlets. This is a simple generalization of our analysis, but allows us to explore the consequence of the non-exactness of the free correlator in $1/c$, and the non-overlapping double-trace operators in more than one channels. Through these explorations we can expect a sharper understanding of the consequence of the hidden conformal symmetry on correlators at one loop level, and hopefully start to understand the true nature of these hidden structures. 

Finally, let us mention another related setup, namely, eleven dimensional supergravity on $AdS_7\times S^4$. The superconformal kinematics for 6d (2,0) theories are quite similar to those for 4d $\mathcal{N}=4$ \cite{Dolan:2004mu,Heslop:2004du,Beem:2015aoa,Rastelli:2017ymc,Zhou:2017zaw,Chester:2018dga}. Although no analogous hidden conformal symmetry exists in eleven dimensions to aid the analysis, the $\langle \mathcal{O}_2^{SG}\mathcal{O}_2^{SG}\mathcal{O}_p^{SG}\mathcal{O}_p^{SG}\rangle$ tree level correlators are nevertheless known for all $p$ from solving a bootstrap problem \cite{Zhou:2017zaw}, where $\mathcal{O}_p^{SG}$ is an one-half BPS operators with conformal dimension $2p$.   Such correlators provide enough input data to compute the $\langle \mathcal{O}_2^{SG}\mathcal{O}_2^{SG}\mathcal{O}_2^{SG}\mathcal{O}_2^{SG}\rangle$ correlator at one loop, by exploiting similar techniques used in this paper. Given that the one-loop Mellin amplitudes are extremely simple for the $AdS_5\times S^5$ background, it is intriguing to see whether such simple behavior is also shared by the one-loop Mellin amplitude in $AdS_7\times S^4$. We can then further take the flat space limit, and study the interplay between CFT and flat space physics in eleven dimensions.

\acknowledgments
Is is a pleasure to thank Agnese Bissi for helpful discussions. This work was partially performed during the 2019 Pollica summer workshop, which was supported in part by the Simons Foundation (Simons Collaboration on the Non-perturbative Bootstrap) and in part by the INFN. XZ also thank the Aspen Center for Physics, supported by National Science Foundation grant PHY-1607611, for providing an excellent working environment during the workshop ``Scattering Amplitudes and the Conformal Bootstrap''. The work of LFA is supported by the
European Research Council (ERC) under the European Union's Horizon 2020 research and
innovation programme (grant agreement No 787185). The work of XZ is supported in part by the Simons Foundation Grant No. 488653.

\appendix

\section{$1/c$-Exactness of $\langle \mathcal{O}_2^{SG}\mathcal{O}_2^{SG}\mathcal{O}_p^{SG}\mathcal{O}_p^{SG}\rangle$ in the Free Theory}\label{apponeovercexact}

\begin{figure}[htbp]
\begin{center}
\includegraphics[width=0.8\textwidth]{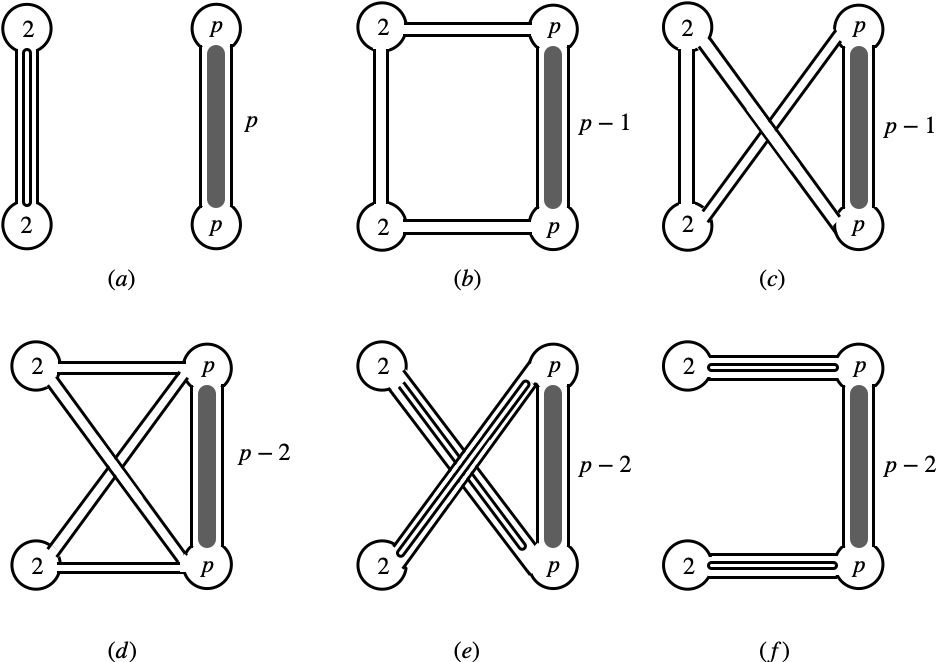}
\caption{Different Wick contractions for $\langle O_2^{SG}O_2^{SG}O_p^{SG}O_p^{SG}\rangle_{free}$. The thick grey line stands for Wick contractions of any number of strands indicated by the number beside it.}
\label{22ppWick}
\end{center}
\end{figure}

In this appendix, we prove that the free correlator $\langle \mathcal{O}_2^{SG}\mathcal{O}_2^{SG}\mathcal{O}_p^{SG}\mathcal{O}_p^{SG}\rangle_{free}$ is $1/c$-exact in the supergravity basis. At zero coupling, the correlator is computed by Wick contractions, and the diagrams fall into six classes (Figure \ref{22ppWick}). We exploit the following simple group theory identity of $SU(N)$ generators  
\begin{equation}\label{TrTrid}
Tr(T^aA)Tr(T^bB)=Tr(AB)-\frac{1}{N}TrATrB\;.
\end{equation}
Note that  for $N$-counting purposes, inserting $O_2^{SG}$ (which has the color structure $Tr(T^aT^b)$) in the Wick contraction diagrams essentially inserts the identity 
\begin{equation}
Tr(T^aT^b)Tr(T^bA)=Tr(T^aA)\;,
\end{equation}
because the second term in (\ref{TrTrid}) vanishes. Therefore  diagram ({\it a\,}) is proportional to 
\begin{equation}
\langle O_2^{SG}O_2^{SG}\rangle\langle O_p^{SG}O_p^{SG}\rangle\;,
\end{equation}
while diagrams ({\it b\,}), ({\it c\,}), ({\it d\,}) become proportional to 
\begin{equation}
\langle O_p^{SG}O_p^{SG}\rangle
\end{equation}
where we have suppressed the spacetime and R-symmetry dependence of the Wick contraction, {\it i.e.}, by setting $\langle X^a(x_1,t_1)X^b(x_2,t_2)\rangle=\delta^{ab}$. On the other hand, diagrams ({\it e\,}) and ({\it f\,}) involve self-contractions, and therefore are not proportional to $\langle O_{p-2}^{SG}O_{p-2}^{SG}\rangle $.\footnote{Recall that $O_p^{SG}$ is generally a mixture of single-trace operators and multi-trace operators for $p\geq 4$.}  However we can prove that these two diagrams always vanish. This follows from the fact that if we cut open the grey line in ({\it e\,}) and ({\it f\,}), each side just corresponds to the OPE of $O_2^{SG}$ and $O_p^{SG}$ to produce a one-half BPS operator with dimension $p-2$. Using the fact that all multi-particle operators are orthogonal to single-particle operators (see (\ref{multiorthosing}) and the discussion around), we  see that the three-point function involving  $O_2^{SG}$, $O_p^{SG}$ and any dimension $p-2$ one-half BPS operator is zero. We normalize the four-point function by dividing it by the $N$-dependence $||\langle O_2^{SG}O_2^{SG}\rangle||$, $||\langle O_p^{SG}O_p^{SG}\rangle||$ of the two-point functions
\begin{equation}
\langle \mathcal{O}_2^{SG}\mathcal{O}_2^{SG}\mathcal{O}_p^{SG}\mathcal{O}_p^{SG}\rangle_{free}=\frac{\langle O_2^{SG}O_2^{SG}O_p^{SG}O_p^{SG}\rangle_{free}}{||\langle O_2^{SG}O_2^{SG}\rangle|| \,||\langle O_p^{SG}O_p^{SG}\rangle||}\;,
\end{equation}
such that the two-point functions have unit coefficients. We find the $N$-dependence only appear as the inverse of $||\langle O_2^{SG}O_2^{SG}\rangle||$ in front of the three nonvanishing connected diagrams, therefore making the four-point function $1/c$-exact.

\section{Some Useful Differential Operators}\label{app:diffops}

In this appendix, we collect useful facts about some differential operators which show up in this paper. We start with the quadratic Casimir operator for $SL(2)$ (or $SU(2)$ by analytic continuation)
\begin{equation}
D_z= z^2\partial_z(1-z)\partial_z-\frac{1}{2}(r+s)z^2\partial_z-\frac{1}{4}rs z\;,\quad r=k_2-k_1\;, \quad s=k_3-k_4\;,
\end{equation}
\begin{equation}
D_\beta= \beta^2\partial_\beta(1-\beta)\partial_\beta+\frac{1}{2}(r+s)\beta^2\partial_\beta-\frac{1}{4}rs \beta\;,\quad r=k_2-k_1\;, \quad s=k_3-k_4\;.
\end{equation}
Using these operators, we can define the $\Delta^{(8)}$ operator \cite{Caron-Huot:2018kta,Aprile:2017qoy,Alday:2017vkk}
\begin{equation}
\Delta^{(8)}_s=\frac{z\bar{z}\beta\bar{\beta}}{(z-\bar{z})(\beta-\bar{\beta})}(D_z-D_\beta)(D_{\bar{z}}-D_\beta)(D_z-D_{\bar{\beta}})(D_{\bar{z}}-D_{\bar{\beta}})\frac{(z-\bar{z})(\beta-\bar{\beta})}{z\bar{z}\beta\bar{\beta}}\;.
\end{equation}
This operator was first defined in \cite{Drummond:2006by} when acting on a function independent of the R-symmetry cross ratios, where it was shown to relate the four-point function of one-half BPS operators $\mathcal{O}_2$ to that of top components of the multiplet. In the hidden symmetry relation (\ref{lls}), it acts on $\lambda_{p_1p_2p_3p_4}\big|_{\log^2 U}$ and gives the leading logarithmic singularity at one loop. It is often convenient to decompose the object which it acts on into different R-symmetry representations labelled by the $SU(4)$ Dynkin label $[m,n-m,m]$
\begin{equation}
F(z,\bar{z};\beta,\bar{\beta})=\sum_{n,m}F_{m,n}(z,\bar{z})Y^{r,s}_{m,n}(\beta,\bar{\beta})\;. 
\end{equation}
Then on each $F_{m,n}(z,\bar{z})$, the differential operators $D$, $\bar{D}$ is replaced by their eigenvalues
\begin{equation}
\Delta^{(8)}F_{mn}Y^{r,s}_{m,n}=\frac{z\bar{z}}{(z-\bar{z})}(D_z-\bar{\mathcal{C}}_{mn})(D_{\bar{z}}-\mathcal{C}_{mn})(D_z-\bar{\mathcal{C}}_{mn})(D_{\bar{z}}-\mathcal{C}_{mn})\frac{(z-\bar{z})}{z\bar{z}}F_{mn}Y^{r,s}_{m,n}
\end{equation}
where 
\begin{equation}
\mathcal{C}_{mn}=\frac{1}{4}(2+m+n)(4+m+n)\;,\quad \bar{\mathcal{C}}_{mn}=\frac{1}{4}(m-n)(m-n-2)\;.
\end{equation}
It is worth pointing out that the operator $\Delta^{(8)}$ is not crossing invariant even when the external weights are equal.

\bibliography{oneloopMellin} 
\bibliographystyle{utphys}

\end{document}